%% file: main.tex
\documentclass[letterpaper,twocolumn,10pt]{article}
\usepackage{comment}
\usepackage{usenix,endnotes}
\usepackage{xcolor}
\usepackage{wrapfig}
\usepackage{graphicx}
\usepackage{tabularx}
\usepackage{arydshln}
\usepackage{multirow, booktabs}
\usepackage{siunitx}
\usepackage{enumerate}
\usepackage{enumitem}
\usepackage{subcaption}
\usepackage{amsmath}
\usepackage{xcolor}
\usepackage[ruled,linesnumbered]{algorithm2e}
\usepackage{listings}
\usepackage{xcolor}
\usepackage[dvipsnames]{xcolor}
\usepackage{amssymb}
\usepackage[normalem]{ulem}
\usepackage{xspace}
\usepackage[available,functional,reproduced]{usenixbadges}

\definecolor{dkgreen}{rgb}{0,0.6,0}
\definecolor{gray}{rgb}{0.5,0.5,0.5}
\definecolor{mauve}{rgb}{0.58,0,0.82}
\definecolor{codegreen}{rgb}{0,0.6,0}
\definecolor{codegray}{rgb}{0.5,0.5,0.5}
\definecolor{codepurple}{rgb}{0.58,0,0.82}
\definecolor{backcolour}{rgb}{0.95,0.95,0.92}

\usepackage{makecell}

\usepackage{framed}

\usepackage{cleveref}
\crefname{section}{\S}{\SS}
\crefformat{section}{\S#2#1#3} 
\crefformat{subsection}{\S#2#1#3}
\crefformat{subsubsection}{\S#2#1#3}

\usepackage{enumitem}
\setlist{itemsep=0pt,parsep=0pt,topsep=0pt}

\usepackage[compact]{titlesec}
\usepackage[font=small,labelfont=bf,aboveskip=8pt]{caption}
\usepackage[all]{nowidow}

\usepackage[most]{tcolorbox}
\definecolor{light-gray}{gray}{0.95}
\newtcolorbox{observation}{
    center,
    width=\linewidth,
    colframe=light-gray,
    colback=light-gray
}

\definecolor{americanrose}{rgb}{1.0, 0.01, 0.24}

\newcommand*\circled[1]{\tikz[baseline=(char.base)]{
            \node[shape=circle,fill,inner sep=1pt] (char) {\textcolor{white}{#1}};}}

\newcommand{\bitx}{\textsf{BitX}\xspace}
\newcommand{\system}{\textsf{ZipLLM}\xspace}

\newcommand{\filededup}{\textsf{FileDedup}\xspace}
\newcommand{\layerdedup}{\textsf{LayerDedup}\xspace} 
\newcommand{\chunkdedup}{\textsf{ChunkDedup}\xspace}
\newcommand{\tensordedup}{\textsf{TensorDedup}\xspace}

\newcommand{\jason}[1]{\textcolor{orange}{Jason: #1}}
\newcommand{\jdel}[1]{\textcolor{orange}{\sout{#1}}}
\newcommand{\zirui}[1]{\noindent\textcolor{blue}{\bf Zirui: #1}}
\newcommand{\yuec}[1]{\noindent\textcolor{red}{Yue: #1}}
\newcommand{\added}[1]{{\color{americanrose}#1}}
\newcommand{\srz}[1]{\noindent\textcolor{purple}{\bf SRZ: #1}}
\newif{\ifrevision}
\revisionfalse

\renewcommand{\jason}[1]{}
\renewcommand{\jdel}[1]{}
\renewcommand{\zirui}[1]{}
\renewcommand{\yuec}[1]{}
\renewcommand{\added}[1]{}
\renewcommand{\srz}[1]{}

\ifrevision

\newcommand{\rdelete}[1]{\textcolor{magenta}{\sout{#1}}}
\else

\newcommand{\rdelete}[1]{}
\fi

\title{{\system}: Efficient LLM Storage via Model-Aware Synergistic Data Deduplication and Compression} 
\author{
Zirui Wang$^1$, Tingfeng Lan$^1$, Zhaoyuan Su$^1$, Juncheng Yang$^2$, Yue Cheng$^1$ \\
$^1$University of Virginia, $^2$Harvard University
}
\date{}

\begin{document}

\maketitle
\pagestyle{empty}

\input{sections/abstract}
\input{sections/introduction}

\input{sections/background}

\input{sections/hf_analysis}

\input{sections/design}
\input{sections/evaluation}
\input{sections/discussion}
\input{sections/conclusion}
\input{sections/acknowledgment}
\bibliographystyle{plain}
\bibliography{reference}

\newpage
\input{sections/appendix}

\end{document}

%% file: sections/abstract.tex
\begin{abstract}

Modern model hubs, such as Hugging Face, store tens of petabytes of LLMs, with fine-tuned variants vastly outnumbering base models and dominating storage consumption. Existing storage reduction techniques---such as deduplication and compression---are either LLM-oblivious or not compatible with each other, limiting data reduction effectiveness. 

Our large-scale characterization study across all publicly available Hugging Face LLM repositories reveals several key insights: (1)~fine-tuned models within the same family exhibit highly structured, sparse parameter differences suitable for delta compression; (2)~bitwise similarity enables LLM family clustering; and (3)~tensor-level deduplication is better aligned with model storage workloads, achieving high data reduction with low metadata overhead.

Building on these insights, we design {\bitx}, an effective, fast, lossless delta compression algorithm that compresses the XORed difference between fine-tuned and base LLMs. We build {\system}, a model storage reduction pipeline that unifies tensor-level deduplication and lossless {\bitx} compression. By synergizing deduplication and compression around LLM family clustering, {\system} reduces model storage consumption by 54\%, over 20\% higher than state-of-the-art deduplication and compression approaches. 

\end{abstract}

%% file: sections/introduction.tex
\section{Introduction}
\label{sec:intro}

Large language models (LLMs) have become foundational tools in modern artificial intelligence (AI). With the rapid progress in open-source LLM development~\cite{meta2024llama3, meta2024llama3-2, meta-llama-3.1-8b, jiang2023mistral7b, meta2025llama4}, millions of LLMs are now publicly available through model hubs such as Hugging Face~\cite{huggingface} and TensorFlow Hub~\cite{tensorflowhub}. These platforms support uploads, downloads, and sharing of base models and fine-tuned variants, enabling users to adapt models to diverse downstream tasks with minimal effort. 

\begin{figure}[t]
\centering
\begin{subfigure}{0.515\linewidth}
  \centering
  \includegraphics[width=\textwidth]{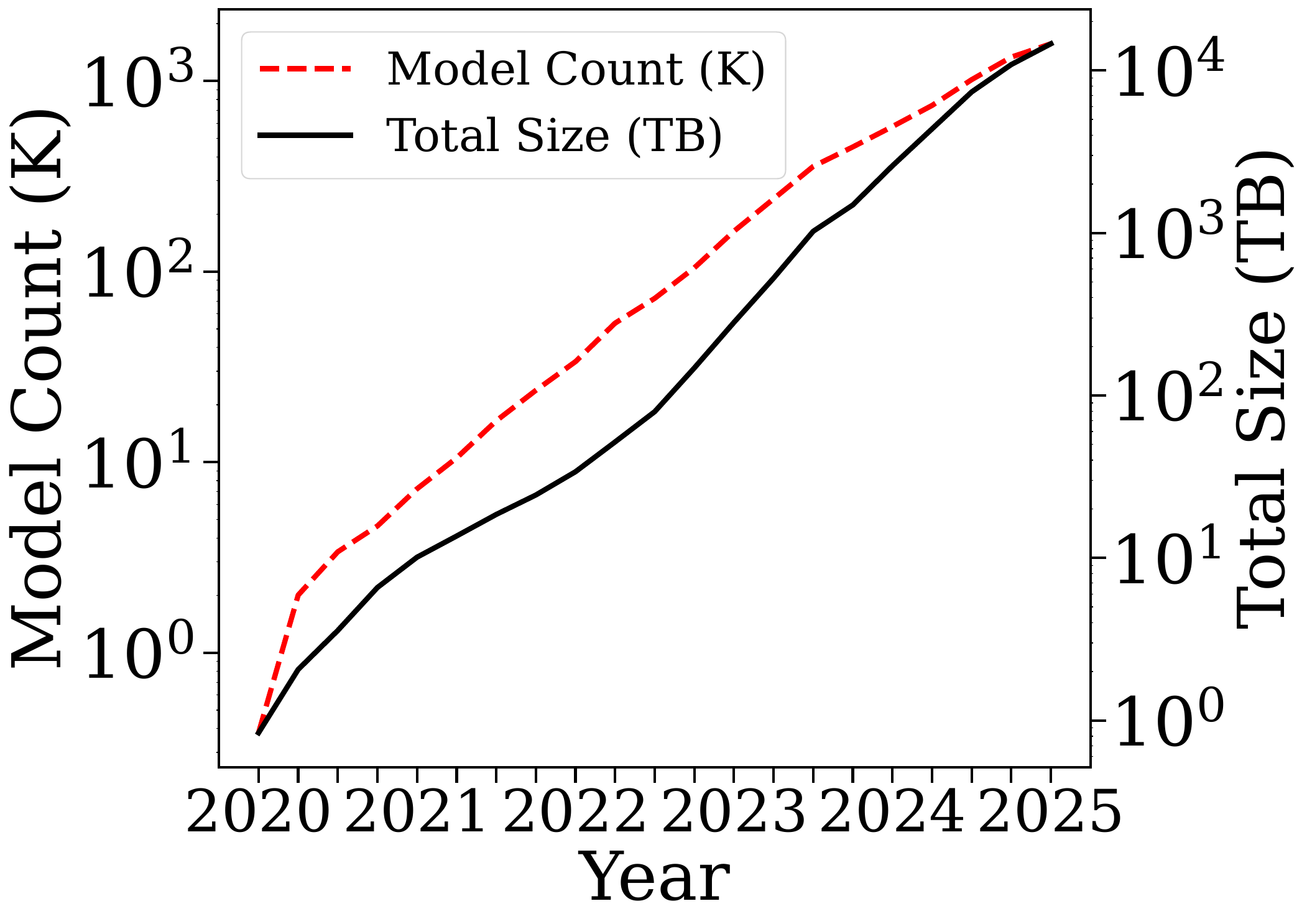}
\end{subfigure}
\hspace{3pt}  
\begin{subfigure}{0.447\linewidth}
  \centering
  \includegraphics[width=\textwidth]{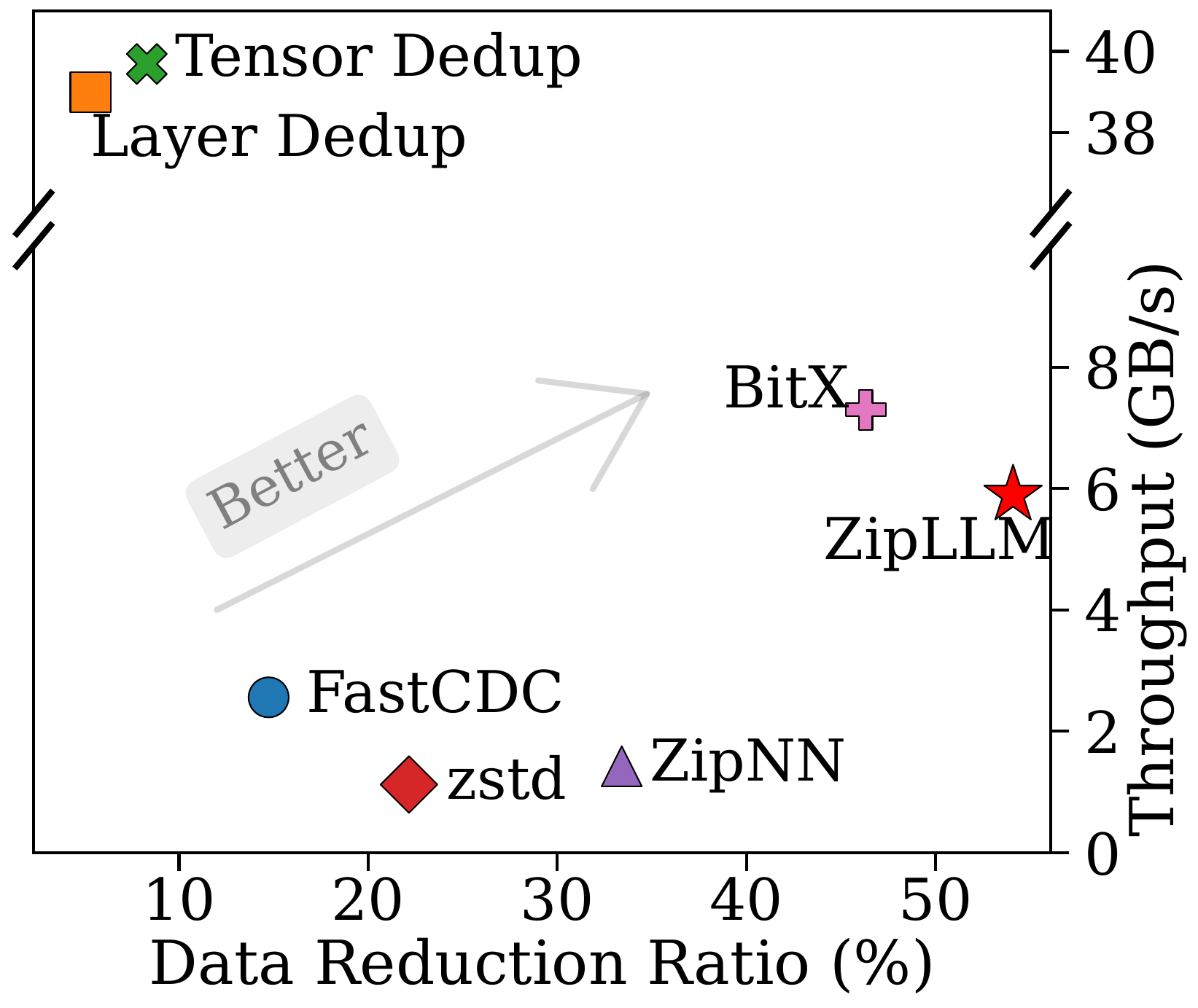}
\end{subfigure}
  \vspace{-14pt}
  \caption{Left: Hugging Face's model count and storage consumption grow exponentially. Right: {\system} achieves both high data reduction and throughput. {\system} represents the end-to-end system throughput. while {\bitx} shows the compression kernel throughput.}
\label{fig:HF_model_size_redcution_ratio}
\end{figure}

This trend has led to an explosion in the number of hosted models. As shown in Figure~\ref{fig:HF_model_size_redcution_ratio}, Hugging Face alone hosts over 14 petabytes (PB) of models (as of Q1 2025), with storage volume growing exponentially, posing a serious threat to the sustainability of machine learning (ML) infrastructure. 

\begin{table*}
\centering
\caption{Comparison of model storage reduction techniques. Note that existing solutions are limited to use either deduplication \textbf{or} compression.}
\vspace{-5pt}
\label{tab:compression_comparison}
\resizebox{\textwidth}{!}{%
\begin{tabular}{@{}lllllll@{}}
\toprule
\textbf{Solution} & \textbf{Compression} & \textbf{Deduplication} & \textbf{Cross-model} & \textbf{Throughput} & \textbf{Storage Reduction} & \textbf{Cons \& Pros} \\ \midrule
\textbf{HuggingFace Xet}~\cite{huggingface-lfs-analysis}
& \textcolor{red}{No} & \textcolor{ForestGreen}{Yes} & \textcolor{ForestGreen}{Yes}
& Low & High & No compression support \\
\textbf{ELF}~\cite{su2024everything}
& \textcolor{ForestGreen}{Yes} & \textcolor{red}{No} & \textcolor{red}{No}
& High & High & Lossy compression \\
\textbf{ZipNN}~\cite{hershcovitch2024zipnn}
& \textcolor{ForestGreen}{Yes} & \textcolor{red}{No} & \textcolor{red}{No}
& Medium & Medium & Ignores cross-model redundancy \\
\textbf{FM-Delta}~\cite{ning2024fm}
& \textcolor{ForestGreen}{Yes} & \textcolor{red}{No} & \textcolor{ForestGreen}{Yes}
& Low & Medium & Requires identical model structure; lacks \texttt{BF16} support \\
\textbf{{\system} (ours)}
& \textcolor{ForestGreen}{Yes} & \textcolor{ForestGreen}{Yes} & \textcolor{ForestGreen}{Yes}
& High & High & Lossless and model structure-aware dedup and compression \\
\bottomrule
\end{tabular}%
}
\vspace{-10pt}
\end{table*}

Two {\it observations} underscore this challenge. 
First, {\it fine-tuned} LLMs vastly outnumber base models and contribute disproportionately to overall storage footprint, despite being only slight modifications. 
Second, LLM storage is dominated by two floating-point formats: \texttt{BF16} and \texttt{FP32}. While \texttt{FP32} is popular in terms of model count (often in smaller models such as those for computer vision), \texttt{BF16} accounts for the majority of total LLM storage size. These trends highlight the need to prioritize LLM-specific storage patterns in future optimizations.

We collect {\it all} public LLM repositories from Hugging Face (Cutoff date March 2025) and conduct a {\it first-of-its-kind, large-scale study focusing on LLM storage}. Our analysis leads to the following {\it insights}:

\begin{itemize}[noitemsep,leftmargin=*]
    \item{\bf Element-wise weight deltas are small and structured within LLM model families.} Fine-tuned models derived from the same base exhibit tiny differences, making them ideal for lossless delta compression. 
    \item{\bf Bitwise similarity enables LLM clustering and lineage tracking.} Bit distance, a new metric that we propose, based on the bitwise Hamming distance, serves as a lightweight, robust signal for identifying LLM families and potentially supporting
    applications like model provenance, duplicate detection, and clustering. 
    \item{\bf Chunk-based deduplication
    is LLM-oblivious and suboptimal for modern model storage.} Chunk-level deduplication, such as content-defined chunking (CDC)~\cite{LBFS, fastcdc, ni2019rapidcdc}, operates on raw byte streams without LLM structure awareness, resulting in the loss of crucial information needed for effective model-aware compression. It also scales poorly with storage capacity.
    \item{\bf Model-aware, tensor-level deduplication is well-suited for LLM-aware lossless compressors}, offering reasonable data reduction, but with significantly higher performance and lower metadata overhead compared to CDC.   
\end{itemize}

This paper makes the following contributions:

\begin{itemize}[noitemsep,leftmargin=*]
  \item We conduct a {\it comprehensive analysis} of Hugging Face's massive-scale model repository, with a focus on how LLM family structure impacts storage redundancy and compression effectiveness.  
  
  \item We introduce a {\it novel metric,  bit distance}, to quantify the similarity between fine-tuned models and their base models. 

  \item Building on this, we design {\bitx}, a {\it highly effective, fast, lossless delta compression algorithm} that compresses LLM variants by encoding XOR-based deltas. 

  \item We identify a {\it new ML system design principle}: for modern model storage systems, deduplication and lossless compression must be co-designed and unified to fully exploit model structure and redundancy.

  \item We build {\system}, 
  a {\it model storage reduction pipeline} that synergizes tensor-level deduplication and lossless {\bitx} compression, achieving higher storage savings for large-scale LLM repositories. 

\end{itemize}

Evaluation results show that {\system} reduces the storage size of 3,048 sampled LLMs by 54.1\%, 20\% higher than the state-of-the-art methods. Meanwhile, {\system} achieves 2$\times$ higher compression throughput (Figure~\ref{fig:HF_model_size_redcution_ratio}). Our implementation is publicly available at: 
    \url{https://github.com/ds2-lab/ZipLLM}.

%% file: sections/background.tex
\sisetup{
  group-separator = {,},
  group-minimum-digits = 4,
  output-decimal-marker = {.},
  detect-all,
  round-mode = places,
  round-precision = 2,
  retain-explicit-plus = false,
  retain-zero-exponent = false,
  zero-decimal-to-integer = true 
}

\section{Background and Related Work}
\label{sec:background}

This section reviews existing techniques for reducing data storage, including both general-purpose and model-aware approaches. A high-level comparison of model storage compression and deduplication methods 
is summarized in Table~\ref{tab:compression_comparison}.

\subsection{Traditional Storage Reduction}

\noindent\textbf{General-purpose Compression.}
General-purpose data compression techniques, such as Zstandard (zstd)~\cite{zstd} and Brotli~\cite{Brotli}, are widely used in storage systems to reduce data size by exploiting local byte-level redundancy through methods like dictionary-based compression~\cite{lz77,lzw} and entropy coding~\cite{huffman,huffman1952method,deutsch1996deflate,duda2013asymmetric,marpe2003context,q_coder_adaptive}. While these methods are data-type agnostic, further compression gains can be achieved when the data type is known. For instance, run-length encoding and delta encoding are highly effective for low-entropy~\cite{RLH} and file changes~\cite{xdelta,macdonald2000file}, and specialized techniques have been developed for columnar and time-series datasets~\cite{kuschewski2023btrblocks,vohra2016practical,liakos2022chimp,blalock2018sprintz,liu2021decomposed,burtscher2007high}. 

Lossy compression methods---such as ZFP~\cite{diffenderfer2019error,lindstrom2014fixed} and SZ~\cite{liang2018efficient,di2016fast,tao2017significantly}---though effective in scientific domains, are unsuitable for model storage due to their inability to guarantee exact recovery. Similarly, while quantization is a popular lossy compression approach for model inference~\cite{lin2024awq,frantar2022gptq,xiao2023smoothquant,zafrir2019q8bert,yao2022zeroquant,dettmers2023spqr}, it is a user-driven choice and orthogonal to storage system design. For model storage, lossless compression that preserves floating-point precision remains essential for correctness. 

\noindent\textbf{General-purpose Deduplication.}
Deduplication is a widely used technique for reducing storage footprint by identifying and storing only unique data blocks, replacing duplicates with references. File-level deduplication, used in systems like Git LFS~\cite{git-lfs}, eliminates exact file copies with minimal metadata overhead but cannot detect partial redundancy. Chunk-level deduplication, especially content-defined chunking (CDC)~\cite{LBFS, fastcdc, TiDedup}, addresses this by splitting files into variable-sized chunks based on content, enabling more effective duplicate detection despite insertions or shifts. 
CDC has been adopted in production systems such as NetApp ONTAP~\cite{netapp-ontap,netapp-tr3966} and Dell EMC Data Domain~\cite{dell-website,dell-datadomain}. 
However, CDC is ill-suited for model storage due to high metadata overhead from massive variable-sized chunks\rdelete{and limited throughput (50-100~MB/s)}, making it impractical for processing large-scale model storage efficiently.

\subsection{Model-aware Storage Reduction}
\label{sec:bg_model_deduplication}

\noindent\textbf{Model-aware Compression.}
State-of-the-art model-aware compression methods exploit model parameter distribution patterns and focus on the IEEE 754 floating-point format~\cite{ieee754standards}, which represents each number using a \textit{sign bit}, an \textit{exponent}, and a \textit{mantissa}. By exploiting the fields with high redundancy, different compression strategies are proposed to reduce model storage space~\cite{su2024everything,hershcovitch2024zipnn,ning2024fm}. 

\textsc{Elf}~\cite{su2024everything} eliminates exponent bits by mapping weights into a normalized range, but it is inherently lossy and unsuitable for model hubs that require exact recovery. ZipNN~\cite{hershcovitch2024zipnn} improves compressibility by reordering float bytes to isolate compressible fields like sign and exponent bits. Although it could exploit cross-model redundancy, the released implementation only supports files with identical sizes, preventing compression when layer dimensions differ (e.g., modified embedding layers).
FM-Delta~\cite{ning2024fm} targets cross-model redundancy by computing weight deltas between fine-tuned and base models. However, it requires strict architectural alignment, lacks \texttt{BF16} support, and achieves low throughput (around 100~MB/s), limiting its applicability at scale. 
DeltaZip~\cite{yao2025deltazip} and BitDelta~\cite{liu2024bitdelta} are two recently proposed techniques that also explore cross-model redundancy. However, both methods are lossy: they approximate or quantize weight differences to reduce GPU memory cost, making them unsuitable for model hubs that require exact recovery of model weights.

Despite being model-aware, \textsc{Elf}, ZipNN, FM-Delta, DeltaZip, and BitDelta require intact model structures (e.g., aligned tensors and parameters) to function effectively. However, once a model is partially deduplicated at the chunk level (e.g., via CDC~\cite{LBFS}), the remaining unmatched regions lose their structural boundaries and appear as fragmented byte sequences. As a result, these residual parts can no longer be processed by compressors that require intact model structure, preventing them from exploiting redundancy.

\noindent\textbf{Model-aware Deduplication.}
Hugging Face employs a two-stage deduplication strategy that combines file-level and chunk-level techniques, including CDC~\cite{huggingface-lfs-analysis}. File-level deduplication removes exact duplicates by comparing content hashes, which is effective for detecting re-uploaded files. To address inefficiencies in handling large files with minor changes---common in fine-tuning and checkpointing---CDC divides files into variable-sized chunks using a rolling hash, enabling only modified chunks to be uploaded. This chunk-based approach, backed by content-addressed storage (CAS), significantly reduces redundancy. Their early findings report up to a 50\% reduction in storage usage and improved upload and download speeds compared to Git LFS~\cite{xet_dedup_blog}.

%% file: sections/hf_analysis.tex
\section{Characterizing Hugging Face Model Storage}
\label{sec:hf_analysis}

The number of public LLM repositories has grown rapidly in recent years, driven by the popularity of open-source model families (e.g., Llama~\cite{zheng2024llamafactory,grattafiori2024llama3herdmodels}, Mistral~\cite{jiang2023mistral7b}) and the widespread practice of fine-tuning base models for domain-specific tasks. This proliferation is further fueled by community sharing and continuous model versioning. As a result, the demand for scalable and efficient storage systems has significantly increased. Public model hubs such as Hugging Face~\cite{huggingface} and TensorFlow Hub~\cite{tensorflowhub} now host millions of LLMs, with the number growing exponentially. This exploding growth places substantial demands on backend storage systems, both in terms of capacity and bandwidth.  
To better understand this landscape, we characterize {\it all} publicly accessible LLMs hosted by Hugging Face, the world’s largest model hub. 

\subsection{Model Storage Explosion}
\label{sec:bg_model_storage_trend}

We define LLM model storage as the storage of parameter files associated with LLMs. Compared to auxiliary files such as configuration files or tokenizers, parameter files dominate the overall storage footprint. These files are typically stored using cloud-based object storage services~\cite{aws_s3,huang2012erasure, s3_interview}, 
which provide efficient access and scalability.  

To quantify this growth, we examine statistics from Hugging Face. As shown in Figure~\ref{fig:HF_model_size_redcution_ratio} (left), the number of public models (including LLMs and non-LLM models) on Hugging Face has surpassed 1.5 million in 2025, up from 500K just one year earlier. The storage footprint grew nearly 6$\times$ during the same period, exceeding 14 PB for public models (excluding private ones) in early 2025. 
Public models are just the tip of the iceberg---a significant portion of models hosted on the platform are private repositories, which are inaccessible to the public. Our observation corroborates with a recent Hugging Face blog~\cite{huggingface-lfs-analysis}. 

\vspace{-6pt}
\begin{tcolorbox}[breakable,width=0.48\textwidth,title={Implications},boxrule=.3mm,colback=white,coltitle=white,left=.5mm, right=.5mm, top=.5mm, bottom=.5mm]    
   \textit{This exponential growth trajectory continues to place mounting pressure on model hubs. Looking ahead, we project that by the end of 2025, model storage demand will continue its exponential trajectory. This data explosion poses a grand challenge to the long-term sustainability of ML/AI infrastructure.}
\end{tcolorbox}

\subsection{Model Storage Format}
\label{sec:bg_model_storage_format}

\begin{figure*}[t]
    \centering
    \begin{subfigure}[t]{0.32\textwidth}
    \includegraphics[width=\linewidth]{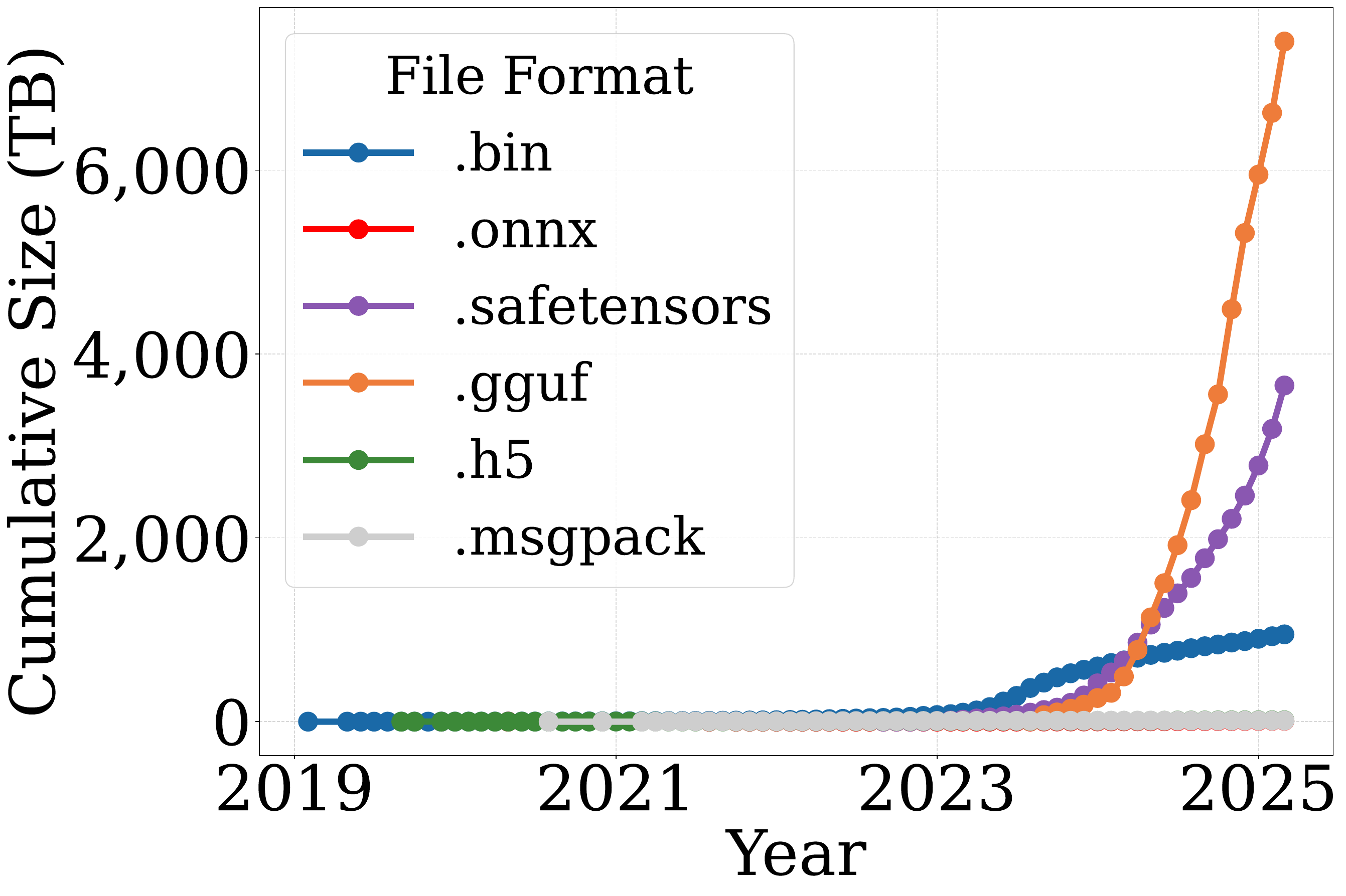}
    \caption{Cumulative model storage by file format.}
    \label{fig:extension_stats_cumulative}
    \vspace{-5pt} 
    \end{subfigure}
    \begin{subfigure}[t]{0.32\textwidth}
    \includegraphics[width=\linewidth]{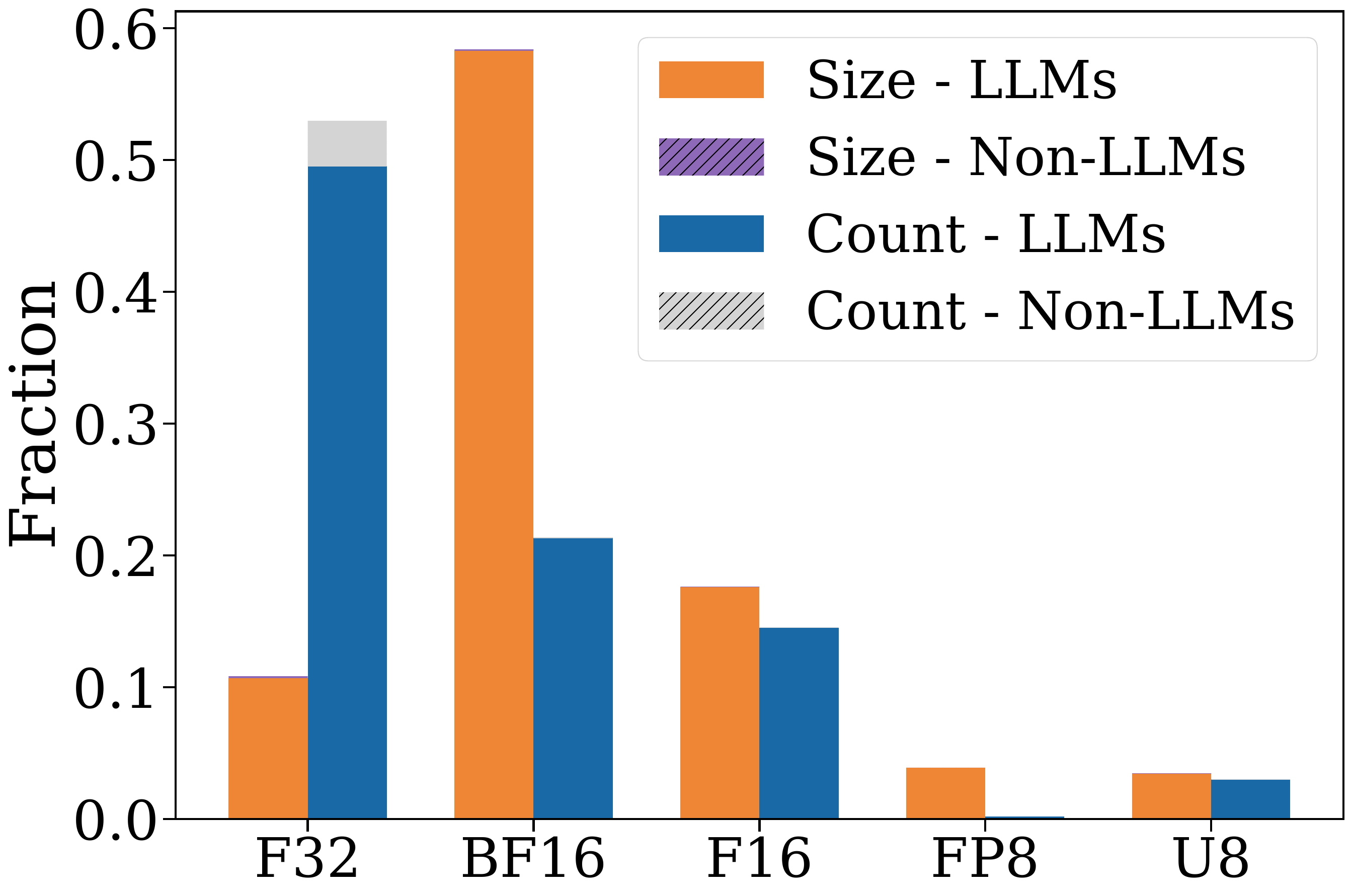}
    \caption{Top data types by size and model count. }
    \label{fig:dtype_size_count}
    \end{subfigure}
    \begin{subfigure}[t]{0.32\textwidth}
    \includegraphics[width=\linewidth]{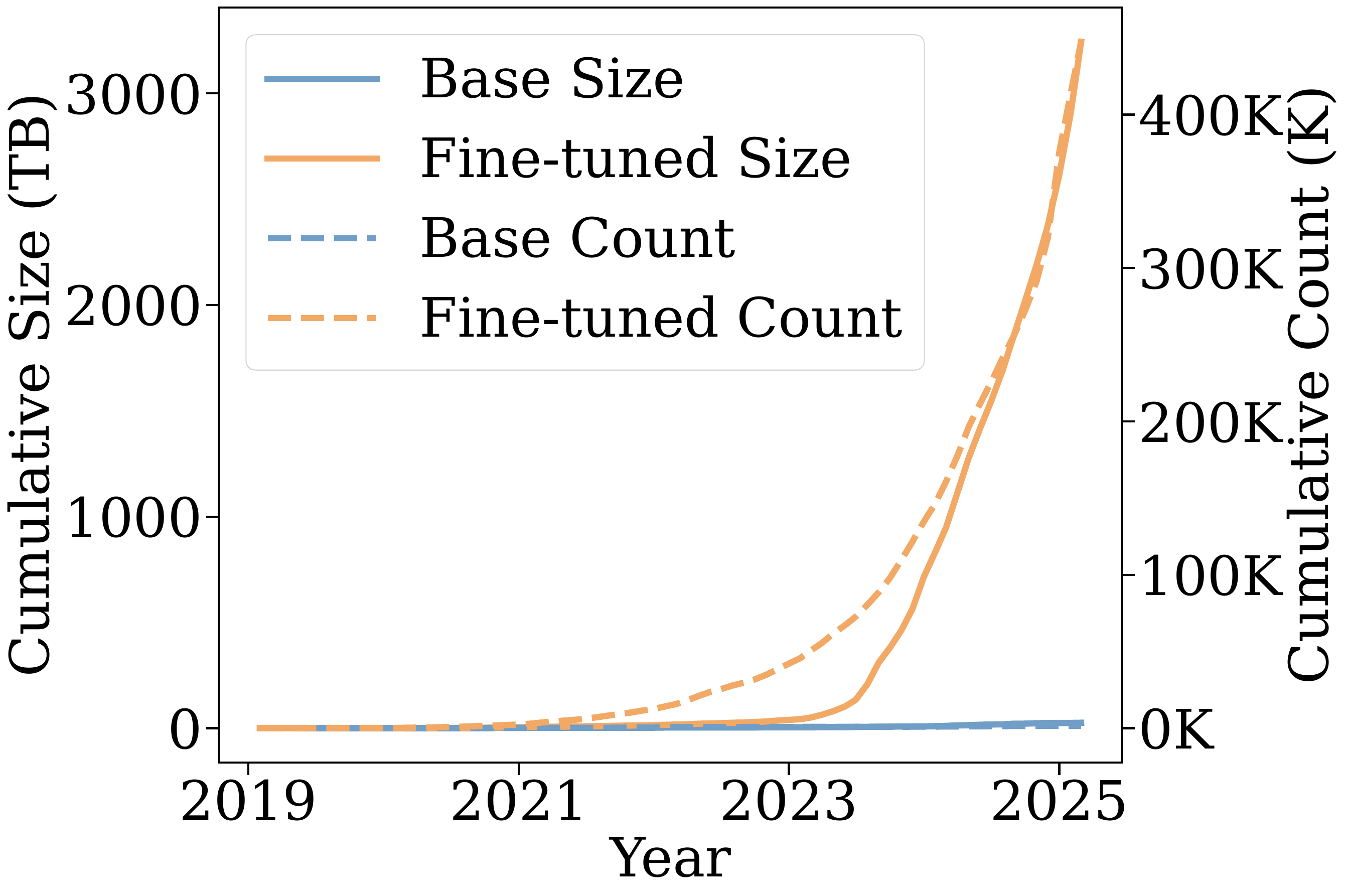}
    \caption{Growth of base and fine-tuned models. }
    \label{fig:comp_base_ft_trend}
    \end{subfigure}
    \vspace{-5pt}
    \caption{Measurement of model repositories on Hugging Face. Note that non-LLMs contribute to a tiny fraction of storage consumption. }
    \label{}
    \vspace{-1em}
\end{figure*}

The shift in model storage formats has significant implications for deduplication and compression. As shown in Figure~\ref{fig:extension_stats_cumulative}, \texttt{safetensors}~\cite{safetensors} and \texttt{GGUF}~\cite{gguf} have become the dominant formats on Hugging Face, together accounting for over 90\% of total storage in 2025.

\noindent\textbf{Safetensors Format.}
\texttt{Safetensors} introduces structured metadata and consistent tensor layouts, which enable model-aware compression and deduplication. Unlike legacy formats like \texttt{.bin} or \texttt{.h5}~\cite{hdf_doc}, which often serialize data with model-specific headers and variable alignment, Safetensors format is also zero-copy and metadata-aware, supporting parallel access to individual tensors without scanning the full file. This enables high-throughput model loading. 

\noindent\textbf{GGUF Format.}
\texttt{GGUF} is a lightweight, extensible format for storing {\it quantized} models, featuring structured metadata for modern tooling compatibility. It addresses limitations of earlier formats and has become the standard for quantized LLMs. 

\vspace{-6pt}
\begin{tcolorbox}[breakable,width=0.48\textwidth,title={Implications},boxrule=.3mm,colback=white,coltitle=white,left=.5mm, right=.5mm, top=.5mm, bottom=.5mm]    
   \textit{These two formats reflect the \textbf{dual} nature of modern LLM workflows: high-precision models are used for training and fine-tuning, while quantized models are used for resource-efficient inference. This dichotomy presents distinct storage patterns and optimization opportunities. As a result, our compression and deduplication techniques are specifically designed to target these formats, which are central to both current and future usage of model storage systems.}
\end{tcolorbox}

\subsection{Data Type Distribution}
\label{sec:bg_model_data_types}

To better understand which floating-point types contribute most to model storage, we analyze the distribution of data types across models. 
As shown in Figure~\ref{fig:dtype_size_count}, \texttt{BF16} is the dominant format in terms of storage consumption, while \texttt{FP32} is the most common in terms of model count.

This discrepancy arises because many models use mixed precision: a few layers in \texttt{FP32} while the majority are \texttt{BF16}~\footnote{In our analysis, if a model contains multiple data types (e.g., both \texttt{FP16} and \texttt{FP32}), it is counted in each corresponding category.}, or they are small non-LLM models (e.g., CV or traditional NLP models) that used \texttt{FP32}.  
As a result, while \texttt{FP32} appears in many models, these models are often small in size. In contrast, \texttt{BF16} is the standard format for large LLM checkpoints, contributing to its substantial share of total storage.

Given these findings, we focus our compression experiments on \texttt{BF16} and \texttt{FP32}, as they are the most prevalent formats in modern repositories. Both use 8-bit exponents, which simplifies our system design and allows unified handling of their binary representations. While we focus our evaluation on \texttt{BF16}, our compression techniques are \textit{data-type-agnostic} and can be generalized to other formats.

\vspace{-6pt}
\begin{tcolorbox}[breakable,width=0.48\textwidth,title={Implications},boxrule=.3mm,colback=white,coltitle=white,left=.5mm, right=.5mm, top=.5mm, bottom=.5mm]
\textit{The vast majority of storage consumption comes from LLMs, particularly those using the \texttt{BF16} format. Non-LLMs contribute minimally in size. As such, future storage optimizations should prioritize LLM-specific formats and patterns to maximize storage efficiency.} \end{tcolorbox}

\subsection{LLM Families} 
\label{sec:comp_family_trend}

\subsubsection{Base and Fine-tuned LLMs}
\label{sec:base_ft_growth}

Beyond data types, another key factor influencing storage footprint is model lineage~\cite{llm_provenance_arxiv25, independent_testing_llm_arxiv25}. It is well known that the majority of today's LLMs are derived from a very small set of base models~\cite{llm_survey_arxiv24, peft_survey_arxiv24}. 
This observation motivates our next analysis, which focuses on LLM families---groups of fine-tuned models that share a common base. Understanding this structure is essential for identifying 
{\it approximate redundancy}\footnote{Defined as hidden, structural redundancy not directly removable via exact deduplication.} 
and guiding targeted storage reduction strategies.

Figure~\ref{fig:comp_base_ft_trend} shows the cumulative count and storage size of base models and fine-tuned models over time. We observe that fine-tuned models have rapidly outpaced base models in both quantity and storage footprint. As of early 2025, fine-tuned models account for at least 3.2\,PB of data, representing 99.22\% of the total storage (3,243.17\,TB out of 3,268.72\,TB). In terms of count, there are 447{,}457 fine-tuned models, comprising 99.64\% of all models. Notably, this is a conservative estimate, as some fine-tuned models lack proper metadata (e.g., model cards) and thus cannot be reliably identified. 

This trend suggests that optimizing the storage of fine-tuned models is crucial to improving the overall efficiency of model repositories. 

\begin{figure}
    \centering
    \includegraphics[width=0.45\textwidth]{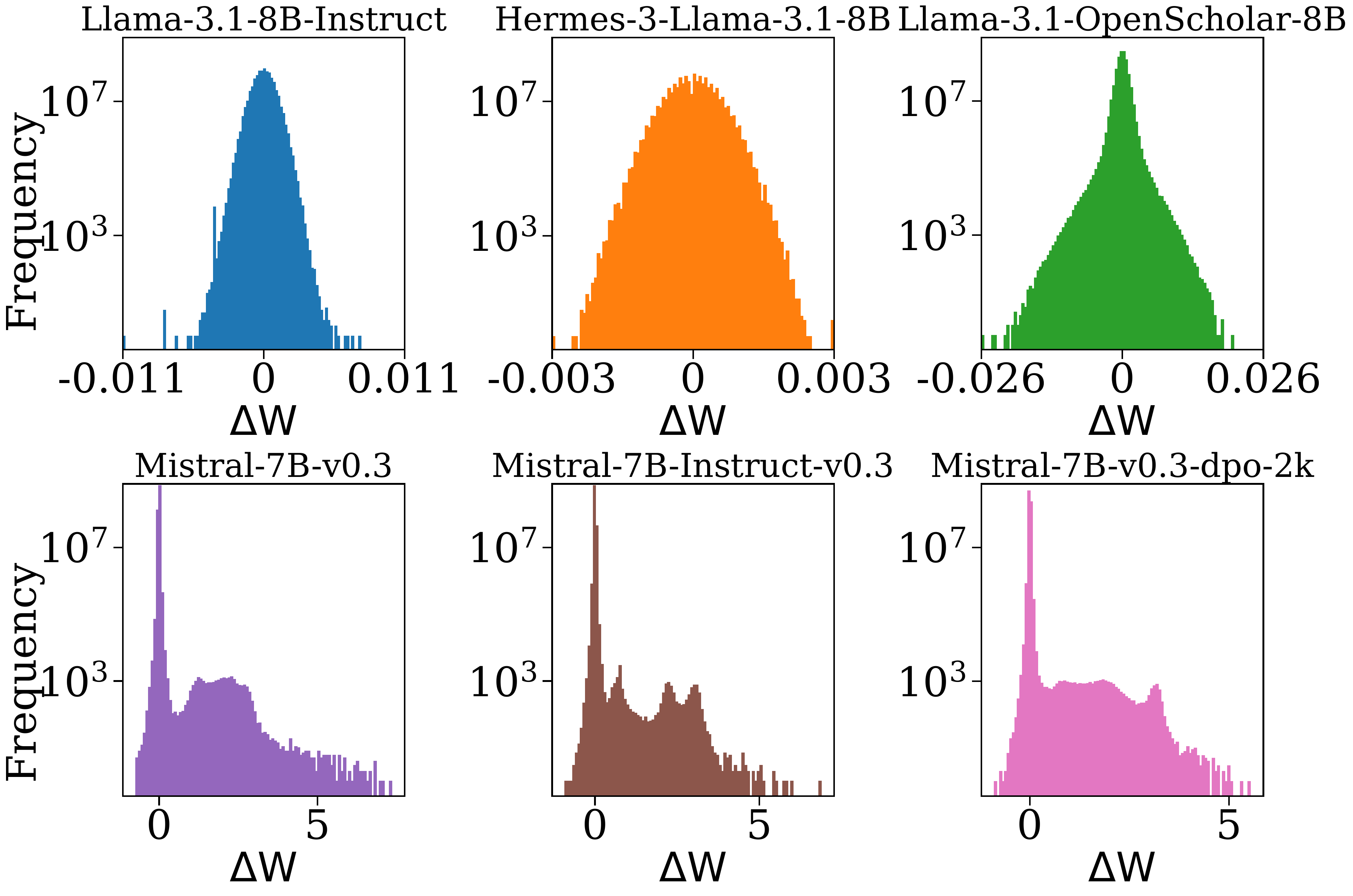}
    \caption{Distribution of element-wise weight differences against \texttt{Llama-3.1-8B}. 
    Top row: Deltas for three models fine-tuned directly from \texttt{Llama-3.1-8B}. 
    Bottom row: Deltas for three models from the \texttt{Mistral-7B-v0.3} family.}
    \label{fig:comp_model_delta}
\end{figure}

\subsubsection{Cross-model Parameter Difference}
\label{sec:comp_model_delta}

Since most fine-tuned LLMs share the same model structure with base models~\cite{grattafiori2024llama3herdmodels, zheng2024llamafactory}---meaning each tensor shares the same shape and position---a natural and direct approach is to analyze the element-wise differences in their weights (model parameters). To verify that such similarity is indeed prevalent, we compute the value differences (delta $\Delta w$) at each parameter position $i$ as $\Delta w_i = w_{i} - \hat{w}_{i} $, where $w_{i}$ and $\hat{w}_{i}$
represent the \( i^{th} \) float value in the fine-tuned and base model, respectively.  
Here, the index \( i \) corresponds to the position of each float in the serialized model file, obtained by traversing all tensors in their original storage order and flattening each tensor in row-major layout. 
This delta is computed across all tensors to capture fine-grained numerical changes between models.

We begin with the \texttt{Llama-3.1-8B}~\cite{meta-llama-3.1-8b} base model and select three of its fine-tuned variants. As shown in Figure~\ref{fig:comp_model_delta}~(top), the delta values are small and centered around zero, with similar bell-shaped distributions across all variants. This indicates that {\it fine-tuned models in the same family introduce only minor modifications to the base model’s weights}.

To validate whether this property holds across model families, we repeat the same analysis using models from the \texttt{Mistral-7B-v0.3}~\cite{jiang2023mistral7b} family. See Figure~\ref{fig:comp_model_delta}~(bottom). Although the architectures are almost identical (except for the embedding and \texttt{lm\_head} layers), the resulting delta distributions are much wider and and asymmetric. 

The element-wise differences are significantly larger, suggesting that models from different origins have less similarity---even if their architecture matches.  

We tested many models and found consistent results, which are omitted due to space limits. 

\vspace{-6pt}
\begin{tcolorbox}[breakable,width=0.48\textwidth,title={Implications},boxrule=.3mm,colback=white,coltitle=white,left=.5mm, right=.5mm, top=.5mm, bottom=.5mm]    
   \textit{
   Element-wise weight deltas can serve as a simple, efficient, and robust tool for identifying model lineage and clustering models by family. Fine-tuned variants derived from the same base model consistently exhibit small and structured deltas, making them well-suited for delta compression. 
   }

\end{tcolorbox}   

\begin{figure}
    \centering
    \includegraphics[width=0.48\textwidth] {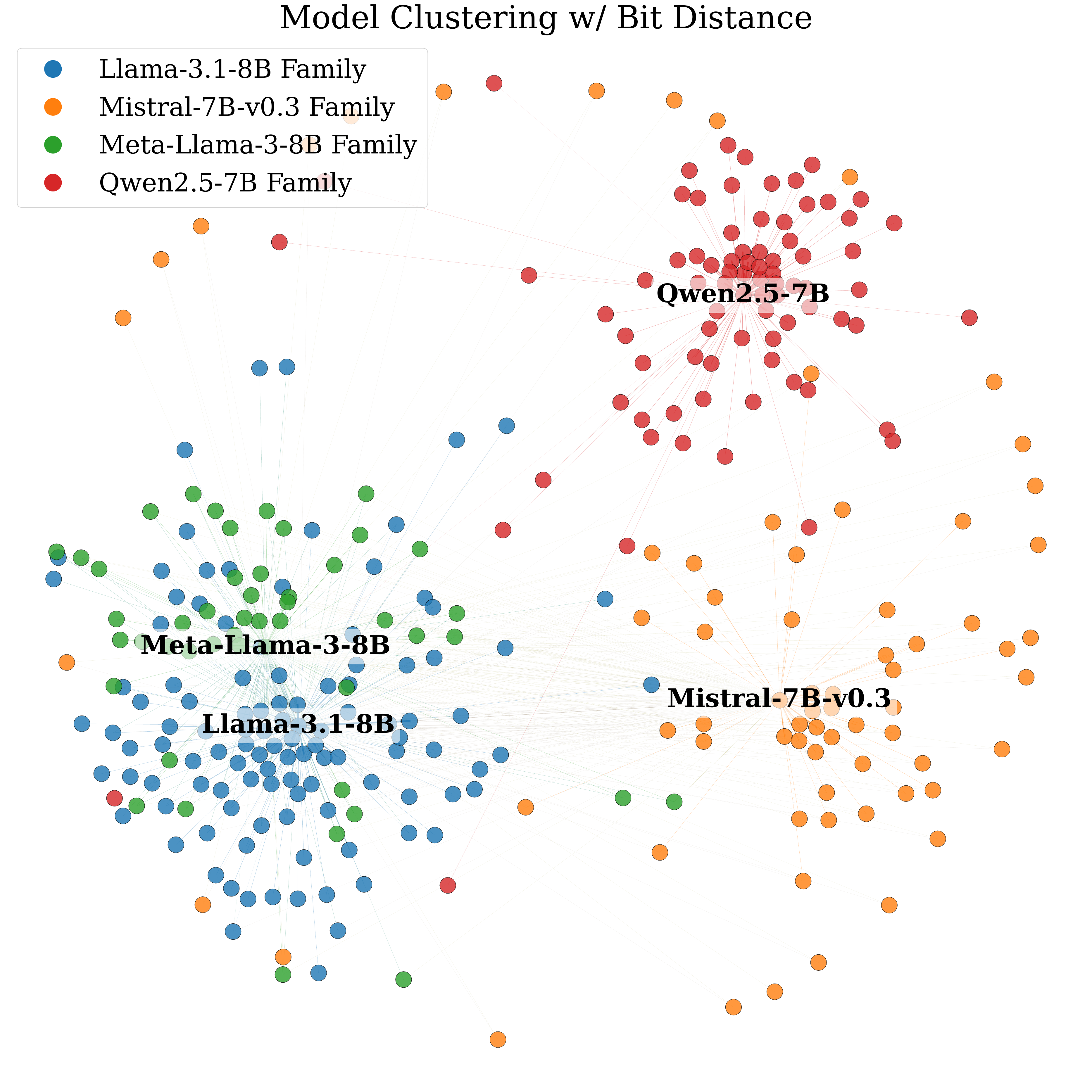}
    \vspace{-20pt}
    \caption{Clustering of 311 LLMs by bit distance.}
    \label{fig:comp_cluster}
\end{figure}

\subsubsection{Cross-model Bit Distance and LLM Clustering} 
\label{sec:comp_cluster}

Building on the observed  correlation between element-wise deltas and model similarity, we propose a bitwise distance metric that measures how many bits differ between two model files. Given two models with the same architecture, we align their floating-point weights in original order and compute the bit distance as follows: 

\vspace{-1.6em} 
\begin{equation}
\texttt{Bit Distance:} \quad 
\mathcal{D}(\mathbf{w}, \mathbf{\hat{w}}) = \frac{1}{n} \sum_{i=1}^n \mathcal{H}(w_i,\hat{w}_i)
\label{eq:bit_distance}
\end{equation}
\vspace{-1.6em} 

\noindent 
Here, \( n \) is the total number of float values in the models. \( w_i \) and \( \hat{w_i} \) denote the \( i\text{th} \) float value from the model pair \( \mathbf{w}\) and \( \hat{\mathbf{w}} \), respectively, both represented in raw binary format. \( \mathcal{H}(w_i,\hat{w_i}) \) computes the number of differing bits (i.e., the Hamming distance~\cite{hamming_distance_bell1950}) between the two binary representations. The final bit distance measures the average number of differing bits per float across the two models. \added{Notably, to avoid bias introduced by popular pruning methods like \cite{frantar2023sparsegpt, sun2023simple}---which may set the same positions to zero across models---we compute the bit distance over non-zero elements only.}  

Using this metric, we compute pairwise distances across 311 models from four major LLM families: \texttt{Llama-3.1}~\cite{meta-llama-3.1-8b}, \texttt{Llama-3}~\cite{grattafiori2024llama3herdmodels}, \texttt{Mistral}~\cite{jiang2023mistral7b}, and \texttt{Qwen2.5}~\cite{yang2024qwen2}. We then construct a similarity graph by connecting model pairs with bit distance below a fixed threshold in Figure~\ref{fig:comp_cluster}.
We observe clear clustering behavior: models within the same family tend to form dense groups, while connections across families are sparse. This supports our earlier hypothesis and findings---models that share a pretrained origin exhibit high structural redundancy, even at the bit level. In contrast, different pretrained base models or models fine-tuned from different bases diverge significantly in their binary representation.

\begin{figure}
    \centering
    \includegraphics[scale=0.125]{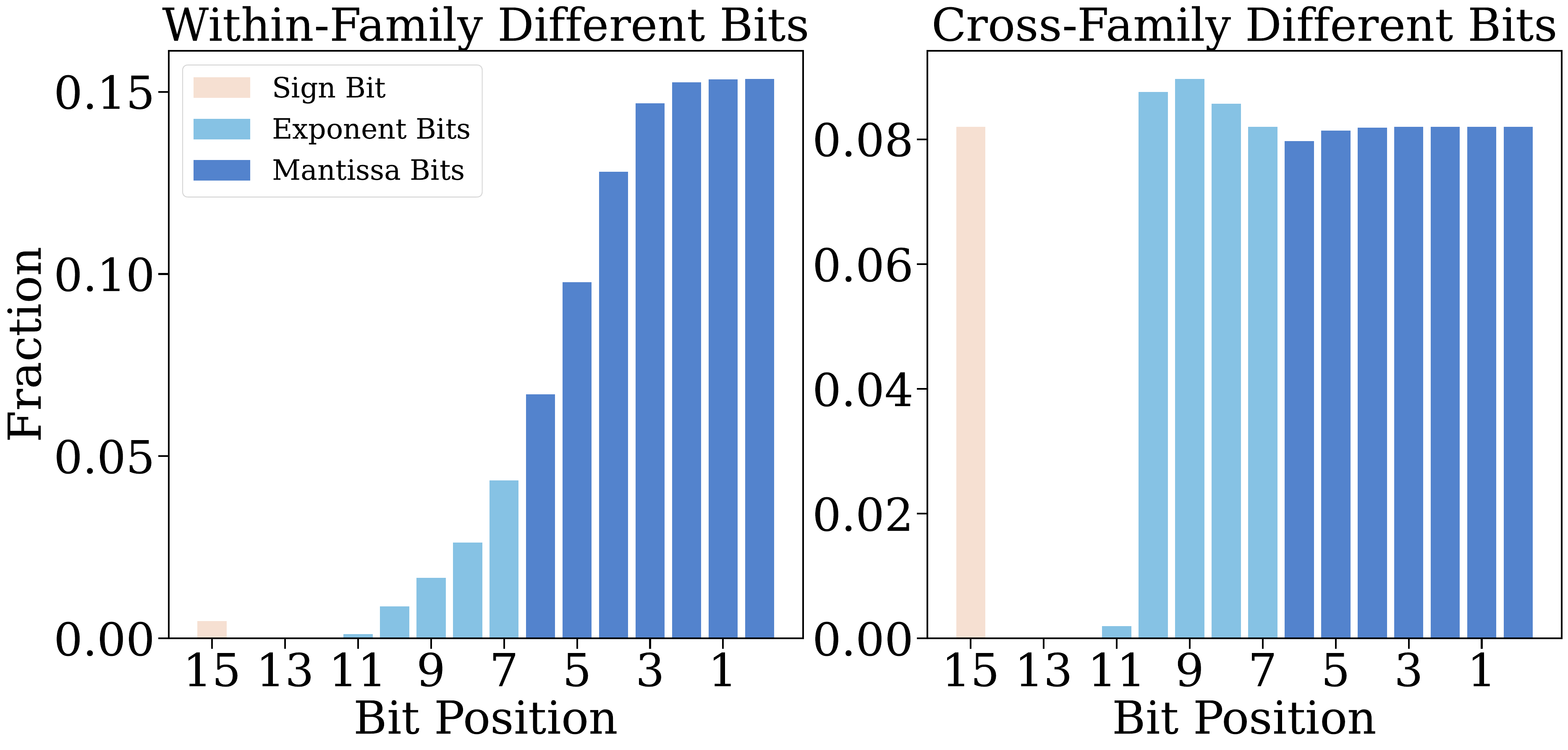}
    \caption{Bitwise contribution breakdown for bit distance. Left: Bit-level differences between a fine-tuned model and its base model within the same LLM family. All models are of \texttt{BF16}. Right: Differences across models from different families. 
    The Y-axis indicates the fraction of total differing bits at each bit position, computed by dividing the number of bitwise XOR results with a 1 at that position by the total number of 1s across all 16 bits.} 
    \label{fig:comp_bit_distance}
\end{figure}

To better understand which bits contribute most to the observed bit-level differences between models, we break down the bit distance by position within the 16-bit \texttt{BF16} format, as shown in Figure~\ref{fig:comp_bit_distance}.
We observe that within the same family, most differences are concentrated in the lower mantissa bits, with the upper mantissa and exponent bits contributing far less, and the sign bit almost never flipping. This indicates a high degree of bit-level similarity, particularly in the high-order bits, which provides a good compression opportunity. In contrast, cross-family comparisons exhibit nearly uniform bit differences across all bit positions, with the exception of a few exponent bits (typically 1–2), which show slightly lower divergence. It reflects their much lower alignment and compatibility for compression. These findings further support the utility of LLM-family-aware compression techniques.

\vspace{-6pt}
\begin{tcolorbox}[breakable,width=0.48\textwidth,title={Implications},boxrule=.3mm,colback=white,coltitle=white,left=.5mm, right=.5mm, top=.5mm, bottom=.5mm]    
   \textit{Bit-level similarity provides a powerful signal for organizing model repositories and guiding LLM storage optimizations. Models that are close in bit distance are more likely to benefit from delta encoding~\cite{wikipedia-vcdiff,bsdiff}, XOR-based compression~\cite{pelkonen2015gorilla}, or structural reuse~\cite{liakos2022chimp}. \\
   Beyond compression, the bit distance metric offers broader implications for large-scale model hubs such as Hugging Face, where accurate and automated identification of model lineage is missing and remains a challenge. Current tools often rely on manually curated metadata. In contrast, bit distance enables content-based provenance analysis, opening the door to a range of applications such as lineage tracking~\cite{llm_provenance_arxiv25}, duplicate detection~\cite{xet_dedup_blog}, model clustering~\cite{hf_model_family_tree_post}, and even LLM testing and evaluation~\cite{independent_testing_llm_arxiv25}. 
   }
\end{tcolorbox}

\subsection{Storage Redundancy in LLMs}
\label{sec:dedup_analysis}
Data deduplication is a widely adopted technique in large-scale storage systems to reduce storage costs by eliminating redundant data~\cite{dedup_tradeoff_fast15,backup_dedup_fast12,dedup_study_fast11, gogetafs_fast25, idedup_fast12}. The effectiveness of deduplication depends heavily on the granularity at which it is applied. As mentioned in \cref{sec:background}, file-level deduplication ({\filededup}) offers low overhead and high throughput, but achieves only limited storage savings. In contrast, chunk-level deduplication ({\chunkdedup}) analyzes data at a finer granularity, enabling greater storage reduction at the cost of significantly higher metadata storage.

\begin{table}[t]
\centering
\caption{{\filededup} stats of Hugging Face model repositories.}
\vspace{-4pt}
\label{tab:file_dedup_stats}
\scalebox{0.85}{
\begin{tabular}{lr}
\toprule
\textbf{Metric} & \textbf{Value} \\
\midrule
Total files & 5,688,779 \\
Duplicate files & 1,182,818 \\
Total size & 11.89 PB \\
Saved size & 0.97 PB (8.2\%) \\
Repos with files can be deduped & 506,337 (33.2\%) \\
\bottomrule
\end{tabular}
} 
\end{table}

\subsubsection{File-level Deduplication}
\label{sec:file_dedup_analysis}
To understand the redundancy landscape in model repositories, we first analyze {\filededup} across all hosted Hugging Face models. 
{\filededup} identifies redundancy by computing cryptographic hashes of entire model files and eliminating duplicates with matching fingerprints. As shown in Table~\ref{tab:file_dedup_stats}, out of 5.6 million model files, approximately 1.18 million files are exact duplicates. {\filededup} can eliminate these duplicates, reducing the total storage footprint by nearly 1~PB. Notably, more than 500,000 repositories---roughly one-third of all currently hosted---contain at least one redundant file, often due to users re-uploading unmodified model artifacts. 

\subsubsection{Deduplication with Content-defined Chunking}
\label{sec:cdc_analysis}
While {\filededup} captures exact matches, we observe significant {\it partial} redundancy between model files that differ slightly, such as checkpoints from the same training run or fine-tuned variants. Deduplication at a finer granularity may help. To this end, we inspect the content of redundant chunks identified by {\chunkdedup}. We find that most deduplicated chunks correspond to serialized tensor data---indicating that the effectiveness of content-defined chunking (CDC)~\cite{fastcdc, ni2019rapidcdc} is largely due to repeated tensors\footnote{An LLM file may contain multiple layers, each with one or more tensors.} across {\it related} models, rather than generic byte-level similarity. This observation reveals that {\it although CDC can detect sub-file redundancy, the underlying source of duplication is often a tensor.} 

\vspace{-6pt}
\begin{tcolorbox}[breakable,width=0.48\textwidth,title={Implications},boxrule=.3mm,colback=white,coltitle=white,left=.5mm, right=.5mm, top=.5mm, bottom=.5mm]    
   \textit{CDC is completely LLM-oblivious and operates directly on raw byte streams. While widely used in industry, it suffers from poor parallelizability and high metadata overhead, limiting its practicality for large-scale model repositories (see \cref{sec:eval_dedup}). By contrast, operating directly at the tensor granularity---where structure is explicitly defined---can achieve similar deduplication ratios but is naturally more parallelizable.}
\end{tcolorbox}

%% file: sections/design.tex
\section{Design}
\label{sec:system_design}

This section presents {\system}, an LLM-aware storage reduction pipeline designed to eliminate redundancy in LLM model hubs efficiently. {\system} combines two complementary strategies—data deduplication and lossless compression—in an LLM-aware manner. These two techniques target different forms of redundancy and operate on separate dimensions, so they do not interfere with each other, but together maximize storage reduction. Unlike conventional CDC-based approaches that operate on raw byte streams, {\system} performs deduplication directly at the tensor level, leveraging model structure explicitly exposed in LLM formats. 
This design not only improves efficiency and deduplication ratio but also preserves the tensor structure required by downstream LLM-aware compressors (e.g., {\bitx}). Built on these principles, {\system} eliminates exact redundancy through {\filededup} and tensor-level deduplication using {\tensordedup}, and reduces approximate redundancy via {\bitx} compression---all while preserving losslessness. We begin by describing each component of {\system} in detail and then explain how they work together in the end-to-end data reduction pipeline.  

\subsection{Model-aware, Tensor-level Deduplication}
\label{sec:llm_dedup_design}
Our earlier analysis from \cref{sec:cdc_analysis} reveals that most repeated chunks correspond to serialized tensors, rather than arbitrary byte patterns. 
This suggests that the underlying source of duplication is {\it structural} in nature---driven by repeated tensors across fine-tuned or checkpointed models.
Instead of relying on expensive compute to find duplicated chunks from long byte streams, {\system} leverages model semantics and performs deduplication directly at the tensor level. 

Modern model formats (\texttt{safetensors} and \texttt{GGUF}) are naturally aligned with {\tensordedup}; they both store model weights in a structured format containing a header followed by serialized tensors. The header contains metadata describing each tensor, including name, shape, data type, and byte offset within the file. By parsing the header first, {\system} can efficiently locate each tensor, enabling parallel processing. 

While we describe deduplication as part of {\system}, it can also be implemented as part of client applications, such as Git LFS~\cite{git-lfs}. When integrated into the client, {\tensordedup} avoids uploading redundant data to the storage server without excessive communication\footnote{{\chunkdedup} is typically performed on storage servers, requiring users to upload full data, as it needs orders of magnitude more hash comparisons.}. This can significantly reduce model upload time and network transfer for users.

\subsection{BitX Delta Compression}
\label{sec:comp_bitX}

Our earlier analysis from \cref{sec:comp_family_trend} reveals a key structural property of modern LLMs: fine-tuned models within the same LLM family exhibit small, consistent differences from their base models, both at the parameter value level (Figure~\ref{fig:comp_model_delta}) and at the bit level (Figure~\ref{fig:comp_cluster}). These differences are often localized and sparse, forming a strong basis for compression. In particular, element-wise deltas show that most parameters remain nearly unchanged during fine-tuning, while the bit-level similarity confirms that related models can be clustered based on shared pretrained origin. Together, these insights motivate a compression strategy that directly encodes the bitwise differences between models. This is the foundation of our Bit-XOR ({\bitx}) approach, which exploits fine-grained redundancy for efficient and lossless model storage reduction.

\begin{figure}
    \centering
    \includegraphics[width=0.45\textwidth]{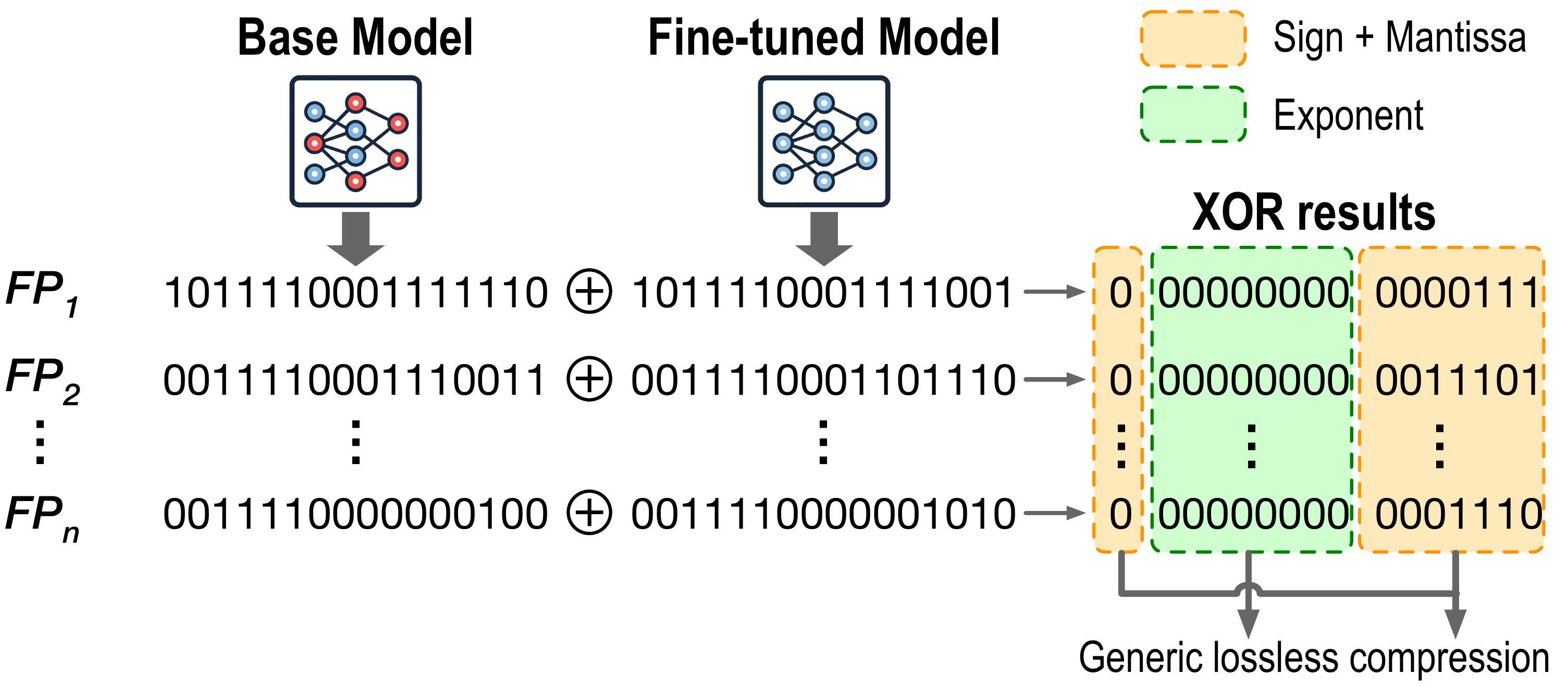}
    \vspace{-5pt}
    \caption{The {\bitx} compression workflow. The example uses \texttt{BF16}, but {\bitx} can support all floating-point types.}
    \label{fig:comp_bitX}
\end{figure}

\noindent\textbf{{\bitx} Workflow.}  
\cref{sec:comp_cluster} shows that models in the same family often exhibit significant bit-level similarity. {\system} introduces a new compression algorithm,  {\bitx}, which exploits this bit-level redundancy to reduce storage consumption.  
Figure~\ref{fig:comp_bitX} illustrates the {\bitx} workflow. 
Given a base model and a fine-tuned model that share the same architecture, {\bitx} first aligns all floating-point values in their original order. For each corresponding pair of floats, {\bitx} performs a bitwise XOR operation. This generates a sequence of XOR results, where many bits---especially in the sign, exponent, and high mantissa bits---are expected to be zero due to high redundancy.
The XOR results capture the fine-grained differences between the models. Since most of the XOR bits are zero, the resulting sequence is highly compressible. {\bitx} then applies a generic lossless compression algorithm, such as zstd~\cite{zstd}, to further reduce storage. This two-stage process efficiently eliminates redundancy by directly encoding only the minimal changes required to reconstruct the fine-tuned model from its base.

\noindent\textbf{Why XOR?}  
We choose XOR rather than numerical differencing because it generates more zero bits at the binary level, leading to higher compressibility. For two similar floating-point numbers, numerical differencing often yields a new value with different exponents and mantissas, making the output denser and harder to compress. 
In contrast, XOR preserves bit-level similarity in the exponent and mantissa, producing sparse outputs that are far more amenable for compression.

By focusing on the bit-level delta between aligned floats, {\bitx} achieves much higher compression ratios (\cref{sec:e2e_eval}) for fine-tuned models than traditional methods, without sacrificing accuracy or requiring any changes to model architectures.

\begin{figure*}[t]
\centering
\includegraphics[width=0.75\textwidth]{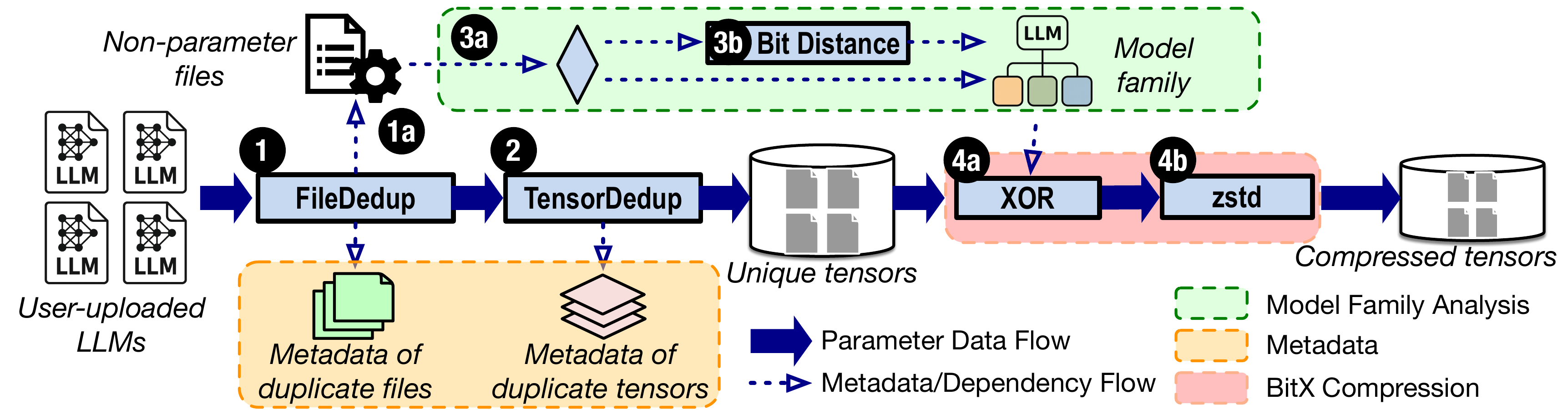}
\vspace{-4pt} 
\caption{Overview of the {\system} storage reduction workflow.}
\label{fig:system_workflow}
\vspace{-12pt} 
\end{figure*}

\subsection{LLM Clustering Thresholding}
\label{sec:clustering_threshold_design}

One key feature of the \bitx workflow is family-based compression, which relies on explicit and accurate family clustering. Missing or inaccurate family information (e.g., pre-labeled tags) can significantly degrade compressibility. To robustly identify LLM models within the same family without pre-labeled metadata, we utilize the proposed bit distance metric (\cref{sec:comp_cluster}) to classify whether two models belong to the same family (within-family) or not (cross-family). First, models with different architectures or tensor shapes can be quickly categorized as cross-family models. For models with the same tensor shapes—which are much harder to distinguish—the decision is made by comparing the pairwise bit distance to a threshold: pairs with distances below the threshold are classified as within-family. In practice, the number of such comparisons can often be reduced to fewer than five, depending on the number of relatively similar variants of base model (e.g., \texttt{Llama-3}, \texttt{Llama-3.1}, \texttt{Llama-3.2}). This section presents our numerical method for determining the clustering threshold. 

Following observations from previous works~\cite{dong2024stbllmbreaking1bitbarrier,ning2024fm}, we assume the parameter weights $\mathbf{w} \sim \mathcal{N}(0, \sigma_w^2)$ and their fine-tuning deviations $\Delta \mathbf{w} \sim \mathcal{N}(0, \sigma_\Delta^2)$ follow symmetric Gaussian distributions centered at 0. Given definition~\eqref{eq:bit_distance}, we define the expected bit distance between the base weights $\mathbf{w}$ and the fine-tuned weights $\hat{\mathbf{w}} = \mathbf{w} + \Delta \mathbf{w}$ as follows:
\vspace{-5pt}
\[
\mathbb{E}[\mathcal{D}(\mathbf{w}, \hat{\mathbf{w}})] = \iint \mathcal{D}(\mathbf{w}, \mathbf{w} + \boldsymbol{\delta}) \, p_{\Delta \mathbf{w}}(\boldsymbol{\delta}) \, p_{\mathbf{w}}(\mathbf{w}) \, d\boldsymbol{\delta} \, d\mathbf{w} 
\]
\noindent 
where $\boldsymbol{\delta}$ denotes the perturbation added during fine-tuning, i.e., $\Delta \mathbf{w}$, and \( p_{\mathbf{w}}(\cdot) \), \( p_{\Delta \mathbf{w}}(\cdot) \) represent the probability density functions of the base weights and their perturbations.

However, the bit distance function $\mathcal{D}(\mathbf{w}, \hat{\mathbf{w}})$ is not continuous. Even small changes in floating-point value can cause sudden bit flips in the bit representation when the $\Delta \mathbf{w}$ cross the ULP (Unit in the Last Place) boundaries, i.e., the smallest spacing between two adjacent representable numbers. For example, changing a value from $1.000$ to $1.001$ may result in $5$ flipped bits in its IEEE 754 representation, even though the numerical change is very small. Because of this, it is challenging to compute the expectation using analytical methods. 
To address this, we adopt a Monte Carlo approach~\cite{metropolis1949monte} to estimate the expected range of bit distance, by sampling from the assumed distributions of the base weights and perturbations: 
\vspace{-8pt}
\[
\hat{\mathbb{E}}[\mathcal{D}(\mathbf{w}, \hat{\mathbf{w}})] = \frac{1}{N} \sum_{i=1}^N \mathcal{D}(\mathbf{w}^{(i)}, \mathbf{w}^{(i)} + \boldsymbol{\delta}^{(i)})
\]
\noindent
where each \( \mathbf{w}^{(i)} \sim \mathcal{N}(0, \sigma_w^2) \) and \( \boldsymbol{\delta}^{(i)} \sim \mathcal{N}(0, \sigma_\Delta^2) \). Here, 
$N$ denotes the number of Monte Carlo samples used to approximate the expectation. We set $N = 100{,}000$ to ensure a stable estimation while keeping the computation efficient.

Empirically, for within-family fine-tuned models, we observe that the base model parameters have a standard deviation in the range of $\sigma_w \in [0.015, 0.05]$, and the fine-tuning deviations lie within $\sigma_\Delta \in [0.00, 0.02]$. The resulting expected bit distance values are consistently within the range of $[3.5, 6]$.

In contrast, cross-family model pairs show bit distance exceeding 6, due to larger weights differences across families as shown in Figure \ref{fig:comp_model_delta}. Notably, for closely related model iterations such as \texttt{Llama-3} and \texttt{Llama-3.1}, the bit distance remains below 6---around 4. Based on these findings, we set a threshold of 4, which yields a classification accuracy of 93.5\% for predicting whether two models belong to the same family. We present detailed sensitivity analysis in \cref{sec:sensi_bit_distance}.

\subsection{Putting It All Together: {\system} Design}
\label{sec:storage_design}

Figure~\ref{fig:system_workflow} illustrates the overall design of our LLM storage reduction pipeline, {\system}, which is tailored to the unique data characteristics of LLM storage. 

In Step~\circled{1}, {\system} deduplicates files by computing content hashes and removing exact duplicates. In Step \circled{2}, {\system} extracts all tensors across repositories and hash them individually to identify repeated tensors. These unique tensors are stored in a global tensor pool. 
{\system} also extracts metadata such as model cards~\cite{hf-model-cards} from non-parameter files (Step \circled{\footnotesize{1a}}), and uses them to group models into families (Step \circled{\footnotesize{3a}}). When the model family metadata is missing or incomplete, {\system} uses bit distance for similarity search (Step \circled{\footnotesize{3b}}) to identify the closest base model (see \cref{sec:comp_cluster}).
\noindent In Step \circled{\footnotesize{4}}, {\system} performs {\bitx} compression, which consists of two sub-steps.
In Step \circled{\footnotesize{4a}}, XOR deltas are computed between fine-tuned tensors and their corresponding base tensors, producing sparse binary differences.
In Step \circled{\footnotesize{4b}}, these XOR results are further compressed using generic algorithms such as zstd, 
yielding the final compact representation.

\subsubsection{File-level Deduplication}
\label{sec:file_dedup_eval}

{\system} first performs {\filededup} on uploaded LLMs. 
This deduplication mechanism is particularly effective for exact reuse cases. For example, we observe that many users upload copies of popular base models (e.g., \texttt{Llama-2-7B}, \texttt{Mistral-7B}) without any modification. These files can be deduplicated entirely without decoding or parsing their contents. 
In addition, {\filededup} is also used as a \textit{prefiltering step} for downstream compression. If a file is unique based on its hash, {\system} proceeds with {\tensordedup} and subsequent compression. If the file is a duplicate, it is simply linked to a previously stored reference to avoid redundant storage.

\subsubsection{Tensor-level Deduplication}
\label{sec:tensor_dedup}

While {\filededup} is effective for detecting exact file reuse, it cannot capture partial redundancy within or across files. To address this, {\system} performs {\tensordedup} by directly operating on tensors stored in model files. 
For each model file, {\system} extracts all individual tensors and compute a hash for each. These hashes are then used to identify duplicate tensors across the entire corpus---within the same file, across multiple files in an LLM repository, or even across different repositories. All \textit{unique} tensors are stored in a global \textit{tensor pool} storage to enable reuse and eliminate redundant storage.  

{\tensordedup} eliminates the inefficiencies of chunk-based deduplication, which suffers from high metadata overhead and limited semantic awareness. 
Compared to layer-level deduplication ({\layerdedup}), which treats an entire layer as one unit, {\tensordedup} offers finer granularity and better tolerance to minor changes in individual layers. 
We compare these deduplication methods later in \cref{sec:eval_dedup} with detailed statistics in Table~\ref{tab:dedup_stats}.

\subsubsection{Lossless Compression}
\label{sec:compression}
After deduplication, {\system} performs LLM-family-aware {\bitx} compression across fine-tuned and base models. 

\noindent\textbf{Model Lineage Extraction.}
This step analyzes the configuration and metadata files extracted from non-parameter files (e.g., \texttt{config.json}, \texttt{README.md}) to identify lineage relationships among models. 
We use a combination of regular expressions and an LLM-based parser to extract base model information. Specifically, we parse architectures, tokenizers, and family identifiers to group structurally similar models.

\noindent\textbf{Bit Distance Matching.}  
If the metadata is missing or incomplete---for example, when the model card only specifies a general base model category (e.g., \texttt{Llama}) without naming a specific base model---{\system} identifies (multiple) likely base models using the structural information. {\system} then computes pairwise bit distances between the fine-tuned model and candidate base models with matching shape and data type (\cref{sec:comp_cluster}). The model with the smallest bit distance is selected as the inferred LLM family. 
\noindent\textbf{Compressing XOR-ed Tensors.}
Once a base model is chosen, {\system} XORs aligned tensors from the fine-tuned and base models. This produces a sparse binary delta, which is lossless while highly compressible. The generated delta will then be compressed by a general-purpose compressor.
By exploiting the approximate redundancy between base and fine-tuned models within the same family, {\system} applies generic compression methods (e.g., zstd) on the resulting deltas, achieving significant storage reduction for vast fine-tuned LLM corpora.

\subsubsection{LLM Serving}
To support efficient LLM loading and serving, {\system} stores minimal metadata alongside compressed model files. For each model, we record its associated base model, the hash of each tensor, the byte offset of each tensor in the original file, and the original \texttt{safetensors} metadata header.
During compression, {\system} additionally stores the base model's tensor hashes used for {\bitx}. At decompression time, {\system} first locates and decompresses each tensor. If a base tensor hash is present, {\system} retrieves the corresponding base tensor and apply XOR to reconstruct the original tensor. All tensors are then reassembled with the metadata header and written in parallel to produce a fully restored model file.  

\noindent\textbf{Fallback Strategy.}
{\system} is designed to be robust even when reference base models are missing. On large model hubs, it is common for multiple copies of the same base model to exist; if one copy is unavailable, {\system} substitutes an equivalent version. In the rare case where all original base models are removed, {\system} selects the most similar fine-tuned model (measured by bit distance) as a surrogate base and computes an additional XOR mask to account for differences. Applying this mask during decompression guarantees exact reconstruction of the target model. For additional robustness, {\system} compares this surrogate-based approach against standalone ZipNN compression and automatically selects the option yielding the better compression ratio.

%% file: sections/evaluation.tex
\section{Evaluation}

\vspace{-4pt}
\subsection{Experimental Setup}

\begin{wraptable}{r}{0.51\columnwidth}
\vspace{-15pt}
\centering
    \caption{Model statistics summary.}
    \label{tab:model_stats}
    \vspace{-5pt} 
    \scalebox{0.85}{
        \begin{tabular}{l r}
            \toprule
            \textbf{Metric} & \textbf{Value} \\
            \midrule
            Model count & \num{3048} \\
            Total size & \num{43.19}~TB \\
            Size after file dedup & \num{41.8}~TB \\
            \bottomrule
        \end{tabular}
    } 
\vspace{-5pt}
\end{wraptable}
\textbf{Dataset.} 
Because of the scale of Hugging Face (tens of PB LLMs stored), we randomly sampled 3,048 open-source LLM repositories from Hugging Face. Our dataset consumes 43.19 TB in raw size (Table~\ref{tab:model_stats}). These repositories span a diverse range of model architectures, including 968 models derived from \texttt{Qwen2.5}~\cite{yang2024qwen2}, 151 from \texttt{Qwen3}~\cite{yang2025qwen3}, 139 from \texttt{Mistral}~\cite{jiang2023mistral7b}, 114 from \texttt{Llama-3}~\cite{meta2024llama3}, 1,431 from \texttt{Llama-3.1}~\cite{meta2024llama3p1}, 47 from \texttt{Llama-3.2}~\cite{meta2024llama3-2}, 135 from \texttt{Gemma-2}~\cite{gemma2024}, and 63 from \texttt{Gemma-3}~\cite{team2025gemma}. 

We exclude LoRA-only repositories from our evaluation. The reason is twofold: (i) LoRA adapters are highly heterogeneous in structure, making them difficult to find a base; and (ii) their sizes are negligible compared to corresponding base models, typically around 1\%. Therefore, after sampling model repositories, we filtered out those that only contain LoRA adapters, leaving us with 3,048 full fine-tuned model repositories. For PEFT-style models, \system by default applies ZipNN to compress the adapters.

\noindent \textbf{Baselines.}
We compare {\system} with both real-world production systems and recent state-of-the-art algorithms: 
\begin{itemize}[noitemsep,leftmargin=*]
    \item{\bf {\filededup} and {\chunkdedup} (FastCDC)} are used by Hugging Face~\cite{xet_dedup_blog}.  
    Because model information is lost during {\chunkdedup}, Hugging Face does not use compression in conjunction with the deduplication. 
    \item{\bf ZipNN} is the state-of-the-art model compressor that groups float numbers' different components
    for compression~\cite{hershcovitch2024zipnn}. Because it does not consider deduplication, we added {\filededup} to ZipNN for a fair comparison. 
    \item{\bf Compress-then-FastCDC} is a baseline we design to study the effect of execution order. 
    In this setting, we first apply a compression algorithm (e.g., zstd) and then perform {\chunkdedup} (FastCDC). 
    This allows us to evaluate how the ordering between compression and deduplication impacts the overall reduction efficiency.

\end{itemize}

Note that we did not compare with FM-Delta~\cite{ning2024fm} and {\textsc{Elf}}~\cite{su2024everything}, because FM-Delta does not support \texttt{BF16}, which is the most popular data type for LLMs, and \textsc{Elf} is lossy. 

At the component level, we compare deduplication across different granularities ({\filededup}, {\layerdedup}, {\chunkdedup}, and {\tensordedup}), as well as zstd compression algorithm.

\noindent\textbf{Implementation.} 
We implemented {\system} entirely in Rust. 
In total, our implementation comprises over 6,000 lines of code. For ZipNN and FastCDC, we use the open-source repos from the authors. 

\noindent\textbf{Metrics.} 
We evaluate system efficacy using multiple metrics.
\begin{itemize}[noitemsep,leftmargin=*]
    \item{\bf Data reduction ratio} calculates the data size reduced by deduplication and/or compression over the original data size. A higher reduction ratio is better. 

    \item{\bf Throughput} measures the speed of deduplication, compression, and decompression of different systems. 

    \item{\bf Scalability} measures how the storage system scales with the number of models. It primarily concerns deduplication, which requires a huge volume of metadata for serving (decompression). Deduplication storage systems are known to suffer from high I/O latency due to excessive metadata overhead~\cite{gogetafs_fast25, idedup_fast12, debnath2010chunkstash}. Therefore, we use metadata size as a proxy for scalability.
\end{itemize}

\noindent\textbf{Setup.} 
We conduct our experiments on an Amazon EC2 \texttt{c6a.48xlarge} instance, equipped with a 96-core AMD EPYC 7R13 processor and 384\,GB of DRAM. All models and associated data are stored on an EBS SSD volume.

\begin{figure}
    \centering
    \includegraphics[width=0.475\textwidth]{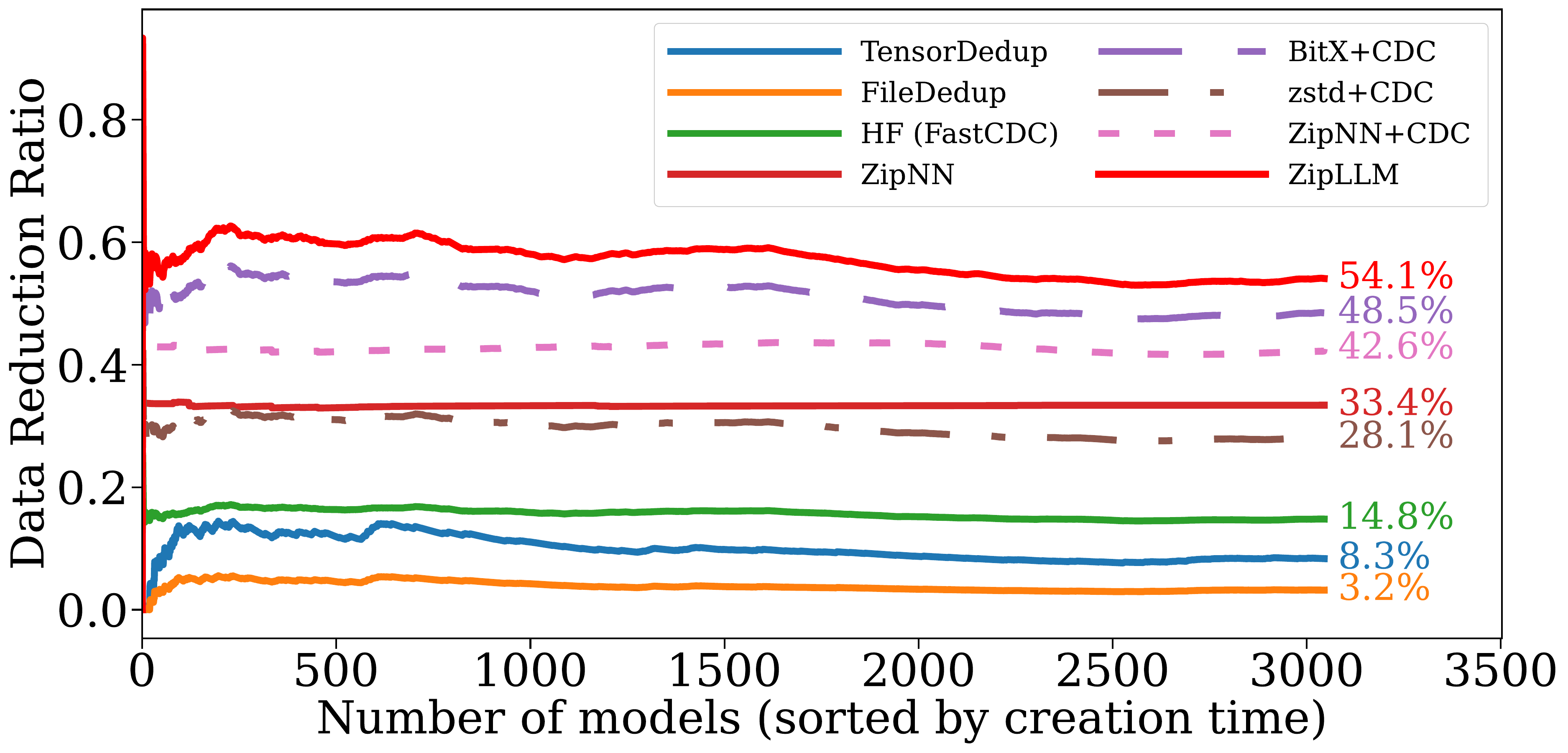}
    \vspace{-15pt}
    \caption{Data reduction ratio vs. model count.}
    \label{fig:e2e_dedup_experiment_result}
    \vspace{-5pt}
\end{figure}

\begin{figure}[t]
    \centering
    \includegraphics[width=0.475\textwidth]{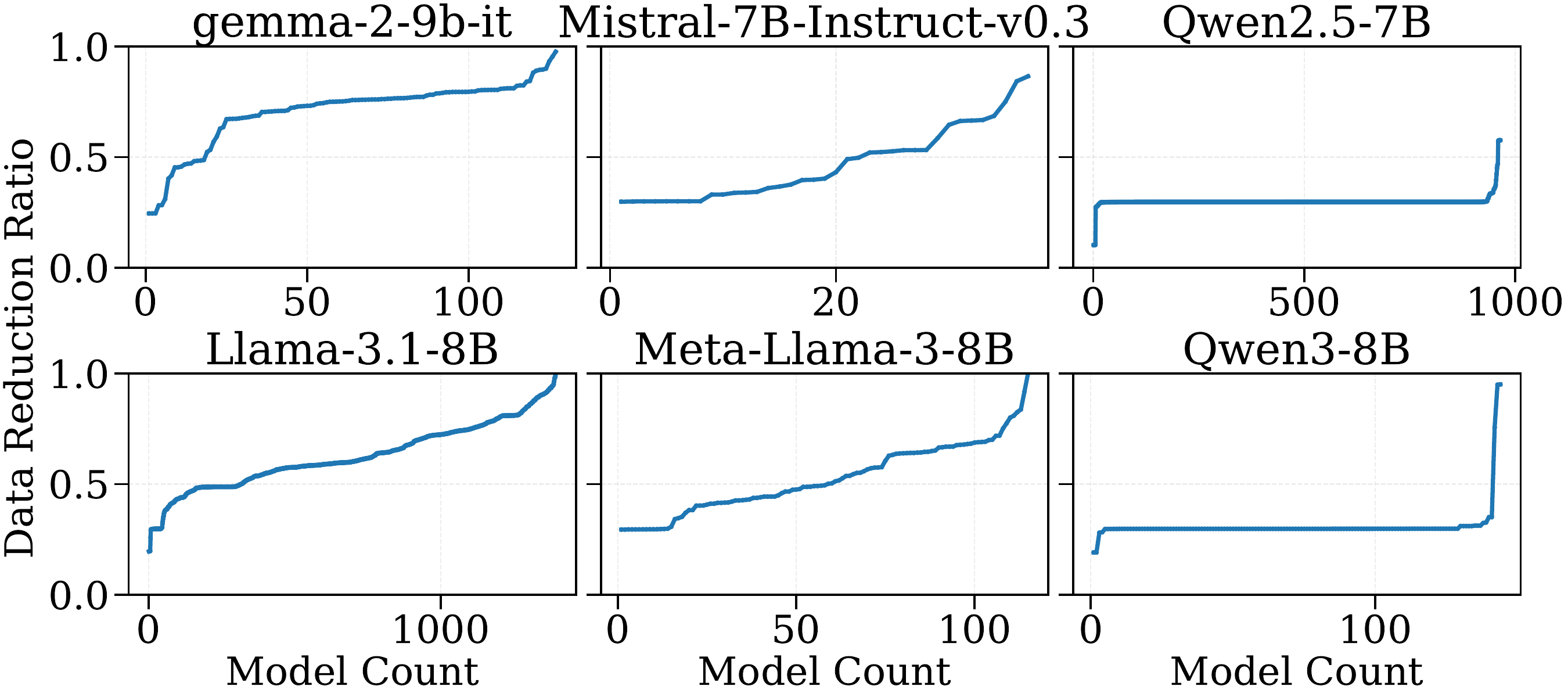}
    \caption{Data reduction ratio (DRR) distributions of six representative base models after applying BitX compression. For each base, all derived fine-tuned models are sorted by DRR in ascending order.}
    \label{fig:ddr_breakdown}
\end{figure}
\subsection{End-to-end Comparison}
\label{sec:e2e_eval}
\subsubsection{Data Reduction Ratio}

To evaluate the overall effectiveness of {\system}, we run the full deduplication and compression pipeline through the entire dataset of 3,048 LLMs.
To simulate the real-world scenario where users continuously upload models to a model hub like Hugging Face, we incrementally increase the number of LLMs and record the corresponding data reduction ratio. 
Figure~\ref{fig:e2e_dedup_experiment_result} shows how different data reduction methods perform as the model storage scales. 
An ideal data reduction method would improve the ratio of redundant data with more uploaded models, leading to higher storage savings. 
The data reduction ratio curve reveals how quickly each method reaches its peak effectiveness and highlights the scalability and long-term benefit of {\system}, which continues to improve and converges later than all baselines.

Figure~\ref{fig:e2e_dedup_experiment_result} shows that {\filededup} achieves limited data reduction,
eventually reducing total storage by 3.2\% on the entire dataset. 
Upon closer examination of the redundant files, we find that most of them happen because the same models are stored multiple times by different users. 
Reducing deduplication granularity often improves the deduplication ratio, as it exposes fine-grained redundancy within files. 
{\chunkdedup} provides up to $4.6\times$ (since $\frac{14.8}{3.2} \approx 4.6$) higher data reduction ratio compared to {\filededup}. However, as we show next in \cref{sec:throughput_eval}, {\chunkdedup} incurs a significant computational and storage overhead, making it difficult to scale. 
ZipNN, a state-of-the-art model-aware compression algorithm, reduces model storage by 33\%, much lower than {\system}. 

The synergy between model-aware deduplication and compression allows {\system} to further boost 
the data reduction ratio to 54.1\%. 
This is because {\system} exploits LLM family information and delta-compresses fine-tuned models. 
{\system} uses \emph{dedup-then-compress}, which outperforms \emph{compress-then-dedup}.
As shown, {\bitx}+FastCDC, ZipNN+FastCDC, and zstd+FastCDC \emph{(compress-then-dedup)} reach 48.5\%, 42.6\%, and 28.1\%, respectively, confirming that compressing first hides redundancy and reduces deduplication effectiveness.

\noindent{\textbf{Per Family Compression Breakdown.}
To better understand how compression effectiveness varies across different model families, we break down the result and present the data reduction ratio distributions of six representative LLM families in Figure~\ref{fig:ddr_breakdown}. We find that {\system} achieves large benefits for most families (e.g., Gemma and Llama-3.1), with median reduction above 0.4-0.7. In contrast, for the Qwen series, the results are more diverse: Qwen includes multiple base variants (e.g., math~\cite{Qwen-math}, coder~\cite{Qwen-coder}, and VL~\cite{Qwen-VL}), and user-provided model cards are often incomplete. These factors make grouping less precise, leading to a DRR close to zstd-level compression. In the future, we plan to improve our grouping algorithm to better handle such heterogeneous families.}

\begin{table}[t]
\small
\centering
\caption{Data ingestion and retrieval throughput with 192 threads.}
\vspace{-8pt}
\label{tab:compression_decompression_throughput}
\begin{tabular}{lcc}
\toprule
\textbf{Method} & \textbf{Ingestion (MB/s)} & \textbf{Retrieval (MB/s)} \\
\midrule
HF (FastCDC) & 2,560     & 9,573 \\
ZipNN        & 1,424    & 9,663 \\
\system      & 5,893  & 7,872 \\
\bottomrule
\end{tabular}
\end{table}

\begin{table*}[htbp]
\small
\centering
\caption{Deduplication statistics. {\tensordedup} strikes a balance
between data reduction ratio and overheads. The projected HF metadata size is  based on 17 PB of models hosted by Hugging Face in 2024~\cite{huggingface-lfs-analysis}, while the estimated metadata size is based on the sampled 3,048 LLMs.}  
\vspace{-5pt} 
\label{tab:dedup_stats}
\begin{tabular}{lrrrrrrr}
\toprule
\textbf{Deduplication Level} & 
\textbf{Unique Hashes} & 
\begin{tabular}[r]{@{}r@{}}\textbf{Avg Size}\\ \textbf{(MB)}\end{tabular} & 
\begin{tabular}[r]{@{}r@{}}\textbf{Max Size}\\ \textbf{(MB)}\end{tabular} & 
\begin{tabular}[r]{@{}r@{}}\textbf{Reduction}\\ \textbf{Ratio}\end{tabular} & 
\begin{tabular}[r]{@{}r@{}}\textbf{Throughput}\\ \textbf{(MB/s)}\end{tabular} & 
\begin{tabular}[r]{@{}r@{}}\textbf{Estimated}\\ \textbf{Metadata (MB)}\end{tabular} & 
\begin{tabular}[r]{@{}r@{}}\textbf{Projected HF}\\ \textbf{Metadata (GB)}\end{tabular} \\ 
\midrule
{\chunkdedup} (FastCDC) & 520,551,953 & 0.087 & 0.25 & 14.8\% & 2560 & 31,772 & 12,505 \\
{\bf \tensordedup (ours)} & 923,384 & 44.9 & 2,000 & 8.3\% & 39,690 & 56 & 22.1 \\
{\layerdedup} & 96,643 & 433.2 & 4,096 & 5.4\% & 38,990 & 5.9 & 2.4 \\
{\filededup} & 12,465 & 3,824 & 15,316 & 3.2\% & 27,099 & 0.76 & 0.30 \\
\bottomrule
\end{tabular}
\vspace{-10pt} 
\end{table*}

\subsubsection{Throughput Performance}
\label{sec:throughput_eval}

\noindent\textbf{Data Ingestion Throughput.}
When the model storage system receives a model upload request, it must perform deduplication and compression before writing the data to the storage. Although the processing can be asynchronous, the data ingestion speed reflects the computational cost required to store models efficiently. 
{\filededup} is the most performant solution due to its simplicity. As a result, its throughput is bottlenecked by the I/O bandwidth. 
HF (FastCDC) achieves 2,560 MB/s (Table~\ref{tab:compression_decompression_throughput}), which is much slower. Moreover, CDC requires sequential boundary detection using a rolling hash, further limiting its scalability.
ZipNN, on the other hand, is the slowest among the three, reaching only 1,424 MB/s (Table~\ref{tab:compression_decompression_throughput}).

Because {\tensordedup} only requires calculating a hash for each tensor, it can scale linearly with the number of tensors. {\system}'s data ingestion throughput only depends on {\bitx} compression. We find that it can achieve a compression speed of over 5,893~MB/s. 

\noindent\textbf{Data Retrieval Throughput.} While computation during data ingestion is performed only once, the computational cost during model downloading---including decompression---must be incurred at each model serving. Since retrieving deduplicated data incurs almost no overhead, we primarily focus on decompression. As a common baseline for decompression, the state-of-the-art generic lossless compressor zstd achieves a single-threaded decompression throughput of 1050~MB/s on our testbed, since its decoding process cannot be parallelized. In contrast, both FastCDC and ZipNN achieve over 9~GB/s (9,573 MB/s and 9,663 MB/s, respectively), while {\system} achieves 7,872~MB/s (Table~\ref{tab:compression_decompression_throughput}). These rates are well above typical disk or network bandwidth, indicating that decompression is not the bottleneck during model retrieval.

\subsection{Breakdown Analysis}
In this section, we break down {\system} into deduplication and compression for detailed benefit analysis.

\subsubsection{Deduplication}
\label{sec:eval_dedup}
To compare different deduplication strategies, we evaluate four methods---{\filededup}, {\layerdedup}, {\tensordedup}, and {\chunkdedup} (FastCDC)---on our 3,048-model dataset. We have discussed {\filededup} in the previous section, so we focus on the others in this section. 

\noindent\textbf{\chunkdedup. } 
As shown in Table~\ref{tab:dedup_stats}, {\chunkdedup} achieves the highest data reduction ratio, removing up to 14.8\% of total data across all models. However, {\chunkdedup} is relatively slow and produces a huge amount of metadata. For example, 520,551,953 unique chunks were produced in our 3,048-model dataset. This results in a vast corpus of metadata to be cached in memory for fast access. Hugging Face stores over 17 PB of models in 2024~\cite{huggingface-lfs-analysis}.
Assuming each chunk requires 64 bytes of metadata~\cite{debnath2010chunkstash}\footnote{This is a reasonable assumption, as metadata typically includes chunk hashes, locations, permissions, reference counts, and timestamps.}, {\chunkdedup} requires over 12.5 TB of storage to just store the metadata. If they were stored in memory with \texttt{c6a.48xlarge} EC2 VMs (384~GB of DRAM)
it would require at least 33 VMs. 
Production systems often replicate metadata for high availability, which would further increase the resource usage. For the same reason, IBM reports that chunk-level deduplication is impossible to deploy at scale~\cite{TiDedup}. Worse, large metadata overhead 
is known to degrade system performance~\cite{debnath2010chunkstash, gogetafs_fast25, idedup_fast12}.  

Another key limitation of CDC-based {\chunkdedup} is its selection of chunk size. To balance between the deduplication ratio and overheads, CDC typically uses large chunk sizes (e.g., 64 KB in Hugging Face~\cite{hf_chunks}). However, model tensors typically range from a few KB to hundreds of MB, meaning that a single tensor often spans multiple chunks, and chunk boundaries may not align with tensor boundaries. This misalignment not only causes boundary-shifting but also complicates the use of downstream model-structure-dependent compressors, which would require extra mechanisms and overhead to recover alignment. In contrast, knowing the data consists of LLMs, we leverage the structural information, such as tensors, to perform more effective, structure-aligned {\tensordedup}.

\noindent\textbf{\tensordedup.}
Because tensors are 100-1000$\times$ larger than chunks, {\tensordedup} produces only 923K unique hashes across the same dataset---a three-order-of-magnitude reduction compared to {\chunkdedup} (Table~\ref{tab:dedup_stats}). 
Meanwhile, it achieves a 
data reduction ratio of 8.3\%. 
This leads to a dramatically smaller metadata index size---approximately 22.1 GB for all the models stored in Hugging Face. 
{\tensordedup} enables more scalable deduplication in terms of memory footprint and system manageability. 
By leveraging explicit tensor boundaries in formats like \texttt{safetensors}, it eliminates the need for rolling hash computations and boundary detection.
Moreover, {\tensordedup} is highly parallelizable as each tensor can be processed independently. 
As a result, {\tensordedup} achieves 15$\times$ higher throughput compared to {\chunkdedup}. 

\noindent \textbf{\layerdedup.}
Besides {\tensordedup}, model-awareness also enables another level of deduplication---{\layerdedup}. 
An LLM typically consists of multiple layers, each of which has multiple tensors. {\layerdedup} performs deduplication at a higher level with coarser granularity, allowing it to generate even fewer entries with less metadata. However, only 5.4\% of the data can be reduced using {\layerdedup}. 

To validate our results, we conduct a direct comparison with the CDC results reported by the Hugging Face Xet team~\cite{xet_dedup_blog} on a public repository~\cite{bartowski-gemma-gguf} used in the blog~\cite{xet_dedup_blog}. 
In Hugging Face production, the 191~GB model was reduced to 97~GB using CDC. 
Our FastCDC baseline and {\tensordedup} both report 93~GB after deduplication. First, this confirms that {\tensordedup} achieves a comparable data reduction ratio. Second, the production {\chunkdedup} result is slightly higher than our experimental results, which we conjecture is due to the production system using a larger chunk size to reduce overhead.

\begin{figure}
    \centering
    \includegraphics[width=0.48\textwidth]
{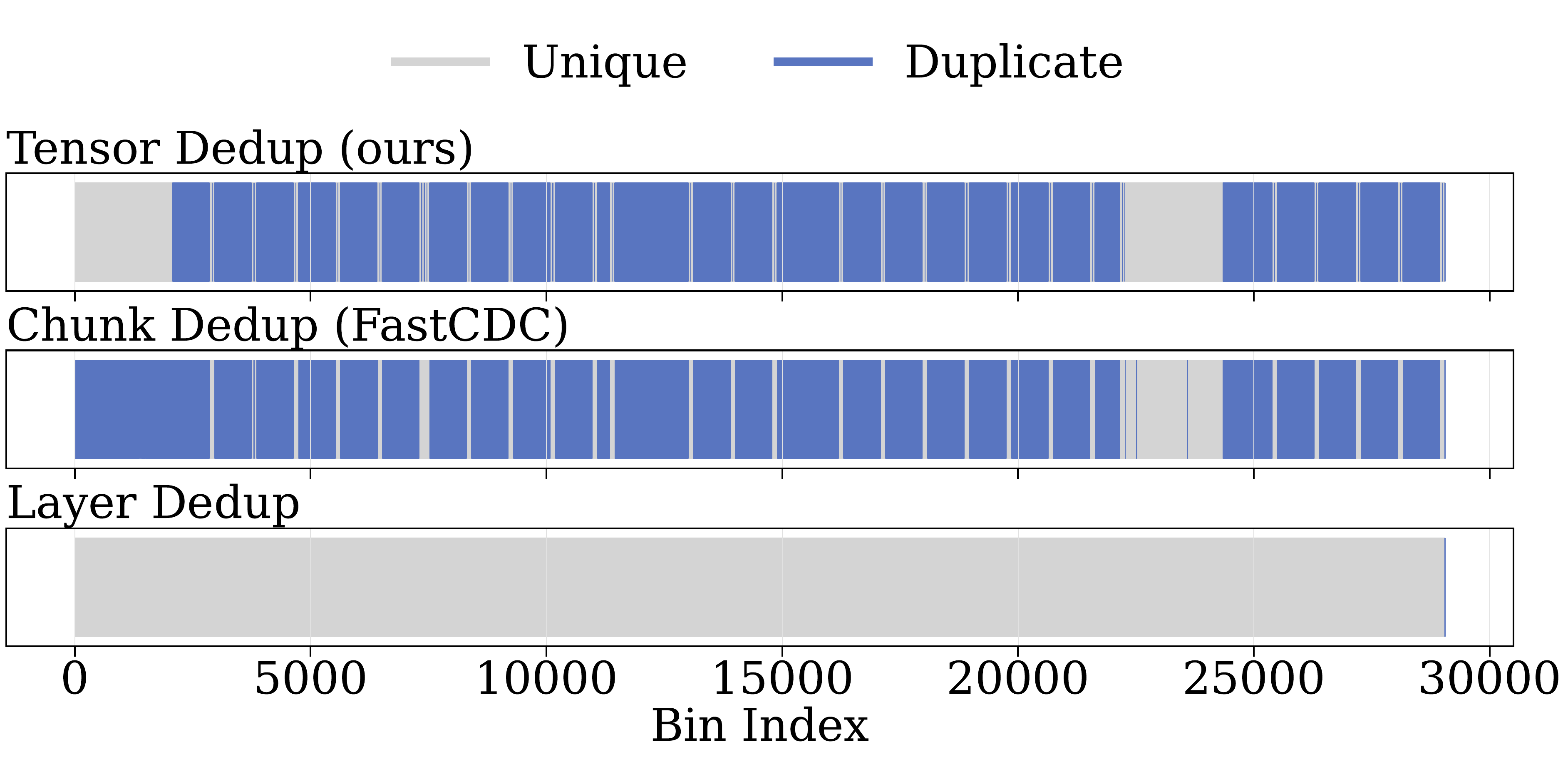}
    \vspace{-15pt} 
    \caption{Visualization of deduplication results under three deduplication levels on a randomly selected LLM repository. Blue indicates duplicate content, while gray indicates unique data. } 
    \label{fig:dedup_compare}
\end{figure}

\noindent\textbf{Visual Comparison.}
We randomly select a model and apply different deduplication methods. 
As shown in Figure~\ref{fig:dedup_compare}, CDC ({\chunkdedup}) and {\tensordedup} produce very similar results. The only major difference appears in the embedding tensor.
This is likely due to vocabulary expansion in fine-tuned models. Although the embedding dimension may change, most of the vocabulary stays the same. Due to its finer granularity, CDC can still match portions of the embedding bytes. In contrast, {\tensordedup} treats each tensor as a whole---any small change makes the entire tensor non-deduplicable. 
For the same reason, {\layerdedup} misses most redundancy because a single modified tensor breaks the whole layer. 
Even so, {\tensordedup} covers nearly all remaining redundancy in the model. It performs as well as CDC, but with significantly less metadata and higher throughput.

\subsubsection{Lossless Compression}
\label{sec:compression_eval}

\begin{figure}[t]
    \centering
    \includegraphics[width=0.475\textwidth]{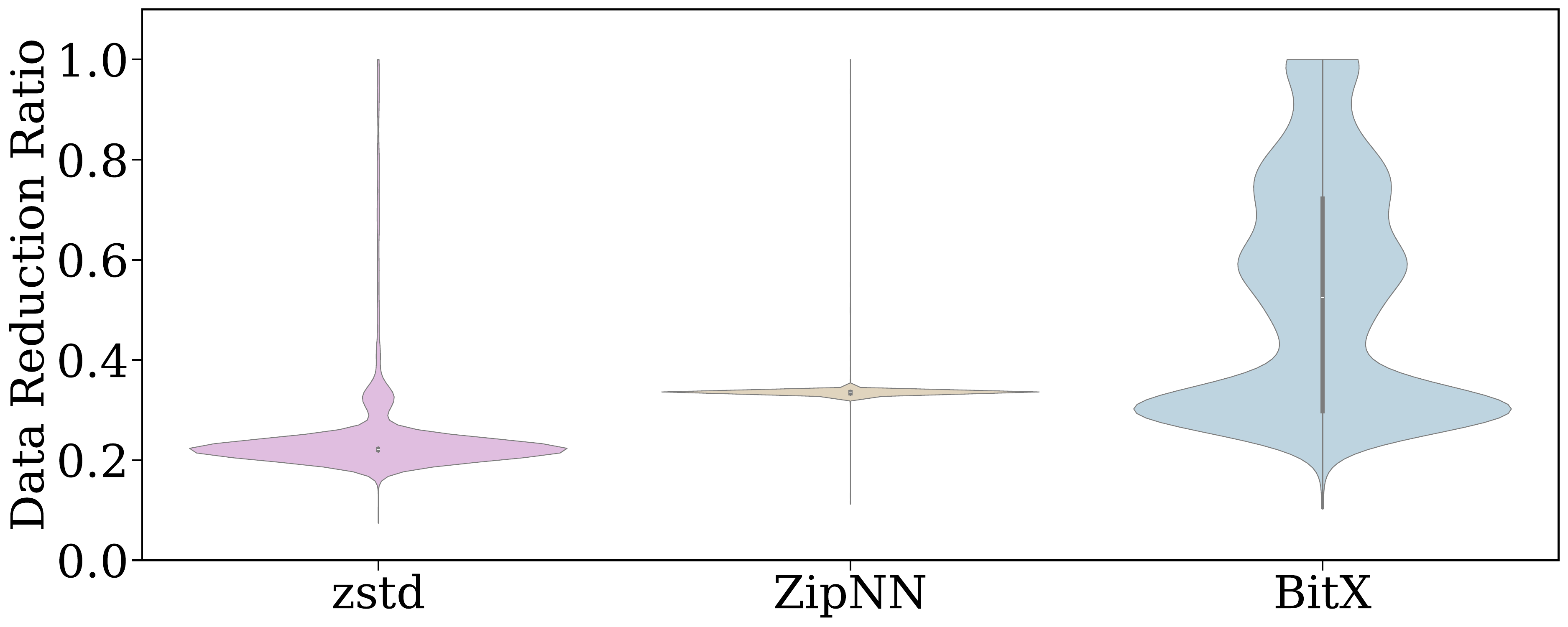}
    \caption{Distribution of data reduction ratio using different compression methods. Each violin plot illustrates the density and spread of data reduction ratios per method, overlaid with a box plot that marks the interquartile range and median.}
    \label{fig:compression_ratio_distribution}
\end{figure}

Figure~\ref{fig:compression_ratio_distribution} shows the data reduction distribution across all 3,048 models using three lossless compression methods: {\bitx} (ours), ZipNN, and zstd. 
Compared to model-oblivious zstd, model-aware compression algorithms, such as ZipNN and {\bitx}, achieve significantly higher data reduction ratios. 
Between ZipNN and {\bitx}, we observe that {\bitx} achieves the best overall data reduction ratio, with many model sizes being reduced by over 50\%.  
The main advantage of {\bitx} over ZipNN is the XOR-based delta compression, which is more effective at compressing models from the same LLM family.  

{\bitx} is both more effective and more performant. 
Because ZipNN groups the sign, exponent, and mantissa from all floating-point numbers in the model for Huffman encoding, it cannot be effectively parallelized, leading to a lower compression throughput as shown in Table~\ref {tab:compression_decompression_throughput}. 
Because {\bitx} operates at the tensor level, each tensor can be processed independently and in parallel, significantly improving compression throughput and achieving over 4$\times$ the throughput of ZipNN.  

These results demonstrate that {\bitx} is both effective and practical for large-scale model storage, achieving significantly better compression while preserving losslessness and model fidelity. 
As previously shown in Table~\ref{tab:compression_decompression_throughput}, {\bitx} also achieves the highest compression throughput, making it well-suited for online model storage reduction at scale. 

%% file: sections/discussion.tex
\section{Discussion}
\label{sec:discussion} 

\noindent\textbf{Encouraging Standardized Tensor Naming and Ordering.}
While \texttt{safetensors} provides a safe and efficient format for model serialization, its specification allows flexibility in tensor naming and does not require preserving the original serialization order. In practice, many models use custom naming conventions or reorder tensors alphabetically by name, which can complicate {\bitx} matching that relies on consistent tensor alignment. We advocate adopting more standardized practices---such as unified naming schemes and optionally recording tensor serialization order---would make it easier to identify corresponding tensor pairs, thereby improving compression effectiveness in systems like {\system}.

\noindent\textbf{Online Quantization and Model Storage Co-design.}  
We observe that many LLM repositories include multiple \texttt{GGUF} files that differ only by quantization method. These variants are often derived from the same base model. This redundancy could be avoided by storing only the base model and the quantization configuration. The backend can then perform online quantization to generate the desired quantized variant on demand, trading additional computation for greater storage savings. {\it We believe these findings open new research avenues in storage-efficient model quantization and encourage further exploration of new ML system design principles that co-designs quantization techniques with storage backends.} 
This approach is not only storage-efficient but also offers flexibility for future quantization schemes without requiring re-uploads of the same base model.

\noindent\textbf{Cost Savings and Practical Impact.}
To understand real-world benefits, we estimate the storage cost reduction achievable by our approach. According to Hugging Face's public statistics, the total model storage footprint reached around 17~PB in 2024~\cite{xet_dedup_blog}. If {\system} achieves a 50\% reduction, this would save approximately 8.5~PB of capacity. Assuming standard Amazon S3 pricing~\cite{aws_s3_pricing}, this translates to an annual cost saving of more than \$2.2M.

%% file: sections/conclusion.tex
\section{Conclusion}
\label{sec:conclusion}

This paper presents {\system}, a model storage reduction pipeline that unifies tensor-level deduplication and a new lossless delta compression called {\bitx} to address the growing scale of LLM storage. 
Our large-scale study reveals key redundancies in LLM repositories and motivates design principles that synergize model storage deduplication with compression. 
{\system} achieves significantly higher storage savings and throughput compared to state-of-the-art approaches, without sacrificing losslessness. 

%% file: sections/acknowledgment.tex
\section*{Acknowledgment}
\label{sec:acknowledgment } 
We thank our shepherd, Dushyanth Narayanan, and the anonymous reviewers for their valuable feedback and comments, which improved the paper. We thank Ajit Banerjee, Di Xiao, and Yucheng Low from the Xet team of Hugging Face for insightful discussions and feedback on this work. This research was supported in part by NSF grants CNS-2322860, OAC-2411009, and OAC-2403313. We also acknowledge support from NSF CloudBank for providing AWS credits, and thank Adobe for their generous research gift.

%% file: sections/appendix.tex
\appendix
\section{Appendix}
\label{sec:appendix}

\begin{figure}[t]
\centering
    \centering
    \includegraphics[width=0.4\textwidth]{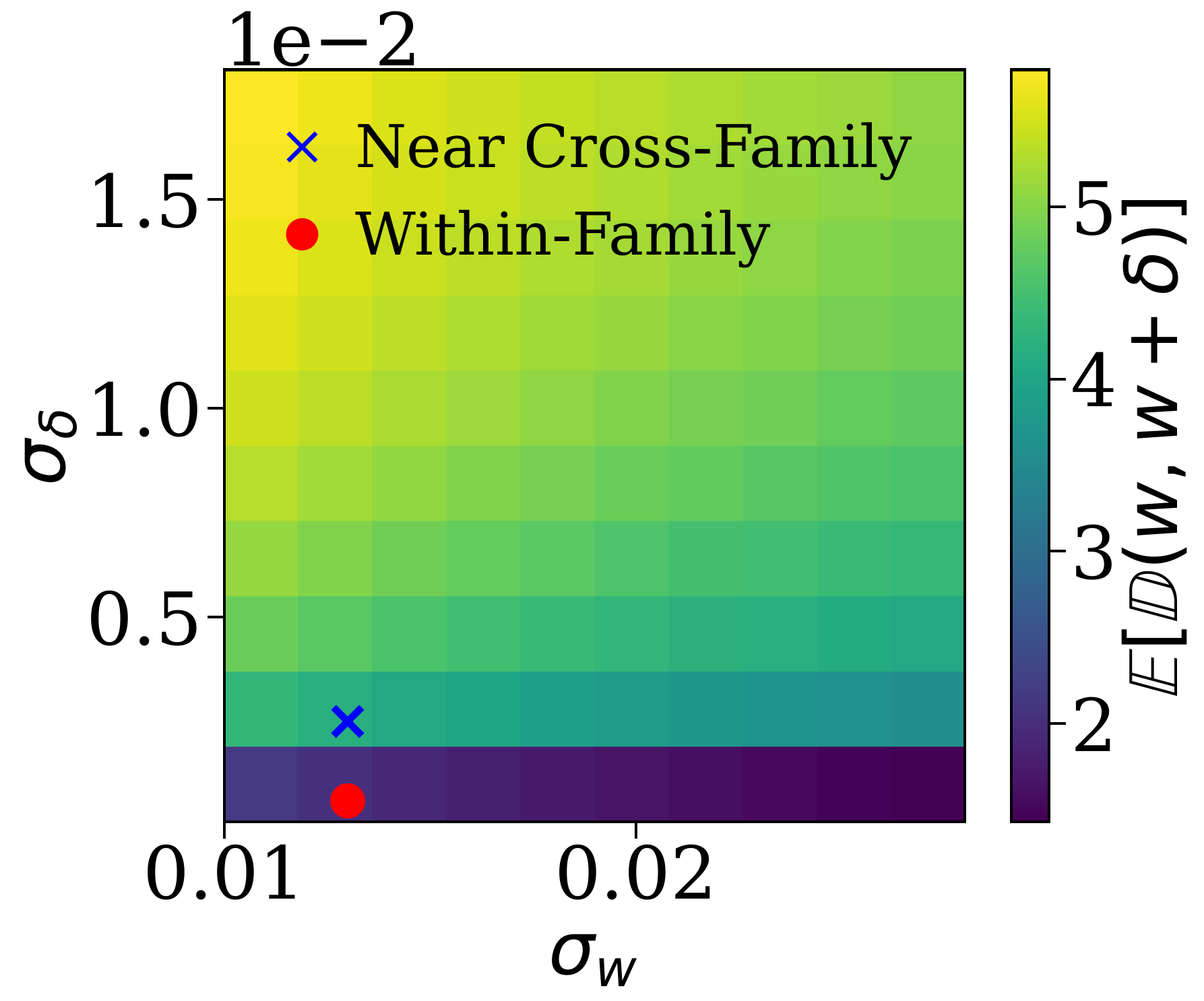}
    \caption{Expected bit distance heatmap. }
    \label{fig:bit_distance_heatmap}
\end{figure}

\subsection{Sensitivity of Clustering Threshold}
\label{sec:sensi_bit_distance}

As discussed in \cref{sec:comp_cluster}, we use a clustering threshold on the bit distance to determine whether a model pair belongs to the same family.

To select a robust and interpretable threshold, we refer to the empirical parameter distribution of popular model families (e.g., \texttt{Llama-3.1}, \texttt{Mistral}, \texttt{Qwen2.5}, etc.), where the standard deviations of base weights and fine-tuning deltas typically fall within $\sigma_w \in [0.01, 0.05]$ and $\sigma_\Delta \in [0.00, 0.02]$ (see Figure~\ref{fig:bit_distance_heatmap}). In this heatmap, darker colors indicate higher expected bit distance values. Under this distribution, the expected bit distance lies within $[1.5, 6]$ based on Monte Carlo sampling. In contrast, cross-family pairs typically exceed 6 due to larger weight deltas, as shown in Figure~\ref{fig:comp_model_delta}.

\begin{figure}[t]
    \centering
    \includegraphics[width=0.37\textwidth]{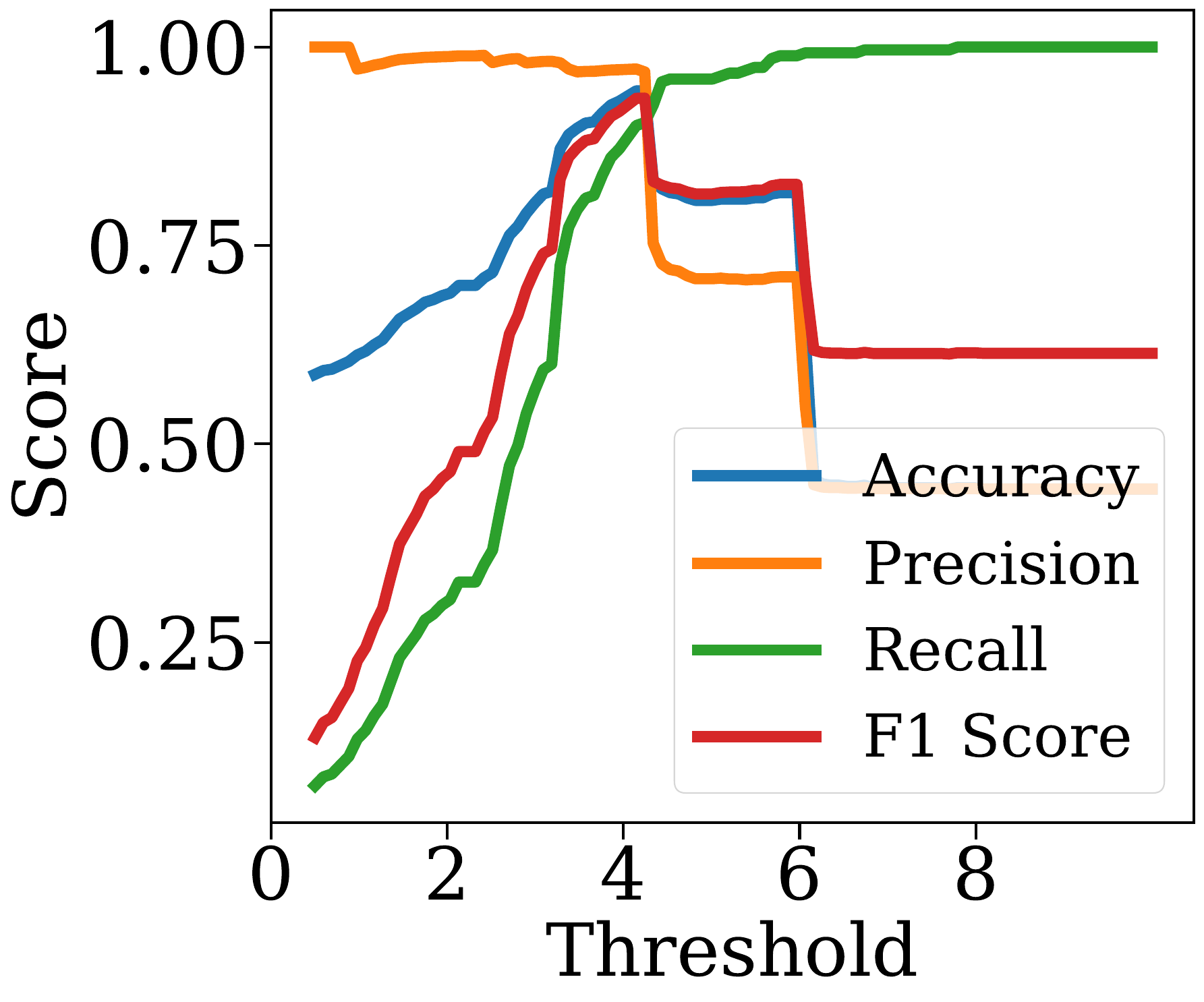}
    \caption{Impact of selected threshold on various metrics.}
    \label{fig:threshold_sensitivity}
\end{figure}

A notable exception is model pairs that are closely related but cross-family (e.g., \texttt{Llama-3} vs. \texttt{Llama-3.1}), which show a lower-than-expected bit distance around 4 (red dot in Figure~\ref{fig:bit_distance_heatmap}). This near-cross-family case reveals a potential risk: although a higher threshold (e.g., 6) would be suggested based on the general distribution, it could lead to false negatives by misclassifying closely related but cross-family models. To address this, we narrow the threshold down to 4, which effectively mitigates such risks. As shown in Figure~\ref{fig:threshold_sensitivity}, the threshold of 4 achieves a high accuracy of \textbf{93.5\%}, while maintaining a good balance across precision (by avoiding misclassifying cross-family models as within-family), recall (since thresholds below 4 would miss many true within-family pairs), and thus results in a strong F1 score.

%% file: reference.bib
@misc{hf_model_family_tree_post,
  title = {{Model Family Tree}},
  howpublished = {\url{https://huggingface.co/posts/mlabonne/611875460328127}}
}

@article{Brotli,
    title	= {Brotli: A General-Purpose Data Compressor},
    author	= {Jyrki Alakuijala and Andrea Farruggia and Paolo Ferragina and Evgenii Kliuchnikov and Robert Obryk and Zoltan Szabadka and Lode Vandevenne},
    year	= {2019},
    URL	= {https://dl.acm.org/citation.cfm?id=3231935},
    journal	= {ACM Transactions on Information Systems}
}

@inproceedings{idedup_fast12,
author = {Srinivasan, Kiran and Bisson, Tim and Goodson, Garth and Voruganti, Kaladhar},
title = {iDedup: latency-aware, inline data deduplication for primary storage},
year = {2012},
publisher = {USENIX Association},
address = {USA},
booktitle = {Proceedings of the 10th USENIX Conference on File and Storage Technologies},
pages = {24},
numpages = {1},
location = {San Jose, CA},
series = {FAST'12}
}

@inproceedings {gogetafs_fast25,
author = {Yanqi Pan and Wen Xia and Erci Xu and Hao Huang and Xiangyu Zou and Shiyi Li},
title = {Don{\textquoteright}t Maintain Twice, It{\textquoteright}s Alright: Merged Metadata Management in Deduplication File System with {GogetaFS}},
booktitle = {23rd USENIX Conference on File and Storage Technologies (FAST 25)},
year = {2025},
isbn = {978-1-939133-45-8},
address = {Santa Clara, CA},
pages = {479--495},
url = {https://www.usenix.org/conference/fast25/presentation/pan},
publisher = {USENIX Association},
month = feb
}

@techreport{zstd,
  title={Zstandard Compression and the application/zstd Media Type},
  author={Collet, Yann and Kucherawy, Murray},
  year={2018}
}

@article{lz77,
  title={A universal algorithm for sequential data compression},
  author={Ziv, Jacob and Lempel, Abraham},
  journal={IEEE Transactions on information theory},
  volume={23},
  number={3},
  pages={337--343},
  year={1977},
  publisher={IEEE}
}

@ARTICLE{lzw,
  author={Welch},
  journal={Computer}, 
  title={A Technique for High-Performance Data Compression}, 
  year={1984},
  volume={17},
  number={6},
  pages={8-19},
  doi={10.1109/MC.1984.1659158}}

@inproceedings{LBFS,
  title={A low-bandwidth network file system},
  author={Muthitacharoen, Athicha and Chen, Benjie and Mazieres, David},
  booktitle={Proceedings of the eighteenth ACM symposium on Operating systems principles},
  pages={174--187},
  year={2001}
}

@inproceedings {TiDedup,
author = {Myoungwon Oh and Sungmin Lee and Samuel Just and Young Jin Yu and Duck-Ho Bae and Sage Weil and Sangyeun Cho and Heon Y. Yeom},
title = {{TiDedup}: A New Distributed Deduplication Architecture for Ceph},
booktitle = {2023 USENIX Annual Technical Conference (USENIX ATC 23)},
year = {2023},
isbn = {978-1-939133-35-9},
address = {Boston, MA},
pages = {117--131},
url = {https://www.usenix.org/conference/atc23/presentation/oh},
publisher = {USENIX Association},
month = jul
}

@ARTICLE{hamming_distance_bell1950,

  author={Hamming, R. W.},

  journal={The Bell System Technical Journal}, 

  title={Error detecting and error correcting codes}, 

  year={1950},

  volume={29},

  number={2},

  pages={147-160},

  doi={10.1002/j.1538-7305.1950.tb00463.x}}

@misc{independent_testing_llm_arxiv25,
      title={Independence Tests for Language Models}, 
      author={Sally Zhu and Ahmed Ahmed and Rohith Kuditipudi and Percy Liang},
      year={2025},
      eprint={2502.12292},
      archivePrefix={arXiv},
      primaryClass={cs.LG},
      howpublished = {\url{https://arxiv.org/abs/2502.12292}}, 
}

@misc{llm_provenance_arxiv25,
      title={Model Provenance Testing for Large Language Models}, 
      author={Ivica Nikolic and Teodora Baluta and Prateek Saxena},
      year={2025},
      eprint={2502.00706},
      archivePrefix={arXiv},
      primaryClass={cs.CR},
      howpublished = {\url{https://arxiv.org/abs/2502.00706}}, 
}

@misc{peft_survey_arxiv24,
      title={Parameter-Efficient Fine-Tuning for Large Models: A Comprehensive Survey}, 
      author={Zeyu Han and Chao Gao and Jinyang Liu and Jeff Zhang and Sai Qian Zhang},
      year={2024},
      eprint={2403.14608},
      archivePrefix={arXiv},
      primaryClass={cs.LG},
      howpublished = {\url{https://arxiv.org/abs/2403.14608}}, 
}

@misc{llm_survey_arxiv24,
      title={Large Language Models: A Survey}, 
      author={Shervin Minaee and Tomas Mikolov and Narjes Nikzad and Meysam Chenaghlu and Richard Socher and Xavier Amatriain and Jianfeng Gao},
      year={2025},
      eprint={2402.06196},
      archivePrefix={arXiv},
      primaryClass={cs.CL},
      howpublished = {\url{https://arxiv.org/abs/2402.06196}}, 
}

@inproceedings{fastcdc,
  title={$\{$FastCDC$\}$: A fast and efficient $\{$Content-Defined$\}$ chunking approach for data deduplication},
  author={Xia, Wen and Zhou, Yukun and Jiang, Hong and Feng, Dan and Hua, Yu and Hu, Yuchong and Liu, Qing and Zhang, Yucheng},
  booktitle={2016 USENIX Annual Technical Conference (USENIX ATC 16)},
  pages={101--114},
  year={2016}
}

@misc{aws_s3_pricing,
  author       = {AWS},
  title        = {{Amazon S3 Pricing}},
  year         = {2025},
  howpublished = {\url{https://aws.amazon.com/s3/pricing/}},
}

@misc{xet_dedup_blog,
  author       = {XetHub},
  title        = {{From Files to Chunks: Improving HF Storage Efficiency}},
  year         = {2024},
  howpublished = {\url{https://xethub.com/blog/from-files-to-chunks-improving-hf-storage-efficiency}},
}

@inproceedings{RLH,
  title={RLH: Bitmap compression technique based on run-length and Huffman encoding},
  author={Stabno, Michal and Wrembel, Robert},
  booktitle={Proceedings of the ACM tenth international workshop on Data warehousing and OLAP},
  pages={41--48},
  year={2007}
}

@article{xdelta,
  title={Efficient algorithms for sorting and synchronization},
  author={Tridgell, Andrew and others},
  year={1999},
  publisher={Australian National University Canberra}
}

@phdthesis{macdonald2000file,
  title={File system support for delta compression},
  author={MacDonald, Josh},
  year={2000},
  school={Masters thesis. Department of Electrical Engineering and Computer Science~…}
}

@article{lin2024awq,
  title={Awq: Activation-aware weight quantization for on-device llm compression and acceleration},
  author={Lin, Ji and Tang, Jiaming and Tang, Haotian and Yang, Shang and Chen, Wei-Ming and Wang, Wei-Chen and Xiao, Guangxuan and Dang, Xingyu and Gan, Chuang and Han, Song},
  journal={Proceedings of Machine Learning and Systems},
  volume={6},
  pages={87--100},
  year={2024}
}

@article{frantar2022gptq,
  title={Gptq: Accurate post-training quantization for generative pre-trained transformers},
  author={Frantar, Elias and Ashkboos, Saleh and Hoefler, Torsten and Alistarh, Dan},
  journal={arXiv preprint arXiv:2210.17323},
  year={2022}
}

@inproceedings{xiao2023smoothquant,
  title={Smoothquant: Accurate and efficient post-training quantization for large language models},
  author={Xiao, Guangxuan and Lin, Ji and Seznec, Mickael and Wu, Hao and Demouth, Julien and Han, Song},
  booktitle={International Conference on Machine Learning},
  pages={38087--38099},
  year={2023},
  organization={PMLR}
}

@inproceedings{ni2019rapidcdc,
  title={RapidCDC: Leveraging duplicate locality to accelerate chunking in CDC-based deduplication systems},
  author={Ni, Fan and Jiang, Song},
  booktitle={Proceedings of the ACM symposium on cloud computing},
  pages={220--232},
  year={2019}
}

@inproceedings{debnath2010chunkstash,
  title={$\{$ChunkStash$\}$: Speeding Up Inline Storage Deduplication Using Flash Memory},
  author={Debnath, Biplob and Sengupta, Sudipta and Li, Jin},
  booktitle={2010 USENIX Annual Technical Conference (USENIX ATC 10)},
  year={2010}
}

@article{su2024everything,
author = {Su, Zhaoyuan and Ahmed, Ammar and Wang, Zirui and Anwar, Ali and Cheng, Yue},
title = {Everything You Always Wanted to Know About Storage Compressibility of Pre-Trained ML Models but Were Afraid to Ask},
year = {2024},
issue_date = {April 2024},
publisher = {VLDB Endowment},
volume = {17},
number = {8},
issn = {2150-8097},
url = {https://doi.org/10.14778/3659437.3659456},
doi = {10.14778/3659437.3659456},
journal = {Proc. VLDB Endow.},
month = apr,
pages = {2036–2049},
numpages = {14}
}

@article{hershcovitch2024zipnn,
  title={Zipnn: Lossless compression for ai models},
  author={Hershcovitch, Moshik and Wood, Andrew and Choshen, Leshem and Girmonsky, Guy and Leibovitz, Roy and Ennmouri, Ilias and Malka, Michal and Chin, Peter and Sundararaman, Swaminathan and Harnik, Danny},
  journal={arXiv preprint arXiv:2411.05239},
  year={2024}
}

@article{ning2024fm,
  title={Fm-delta: Lossless compression for storing massive fine-tuned foundation models},
  author={Ning, Wanyi and Wang, Jingyu and Qi, Qi and Zhu, Mengde and Sun, Haifeng and Cheng, Daixuan and Liao, Jianxin and Zhang, Ce},
  journal={Advances in Neural Information Processing Systems},
  volume={37},
  pages={66796--66825},
  year={2024}
}

@misc{ieee754standards,
  title        = {{IEEE 754-2019: IEEE Standard for Floating-Point Arithmetic}},
  howpublished = {\url{https://standards.ieee.org/ieee/754/6210/}},
}

@misc{grattafiori2024llama3herdmodels,
      title={The Llama 3 Herd of Models}, 
      author={Aaron Grattafiori and Abhimanyu Dubey and Abhinav Jauhri and Abhinav Pandey and Abhishek Kadian and Ahmad Al-Dahle et al.},
      year={2024},
      eprint={2407.21783},
      archivePrefix={arXiv},
      primaryClass={cs.AI},
      howpublished = {\url{https://arxiv.org/abs/2407.21783}}, 
}

@misc{jiang2023mistral7b,
      title={Mistral 7B}, 
      author={Albert Q. Jiang and Alexandre Sablayrolles and Arthur Mensch and Chris Bamford and Devendra Singh Chaplot and Diego de las Casas and Florian Bressand and Gianna Lengyel and Guillaume Lample and Lucile Saulnier and Lélio Renard Lavaud and Marie-Anne Lachaux and Pierre Stock and Teven Le Scao and Thibaut Lavril and Thomas Wang and Timothée Lacroix and William El Sayed},
      year={2023},
      eprint={2310.06825},
      archivePrefix={arXiv},
      primaryClass={cs.CL},
      howpublished = {\url{https://arxiv.org/abs/2310.06825}}, 
}

@inproceedings{zheng2024llamafactory,
  title={LlamaFactory: Unified Efficient Fine-Tuning of 100+ Language Models},
  author={Yaowei Zheng and Richong Zhang and Junhao Zhang and Yanhan Ye and Zheyan Luo and Zhangchi Feng and Yongqiang Ma},
  booktitle={Proceedings of the 62nd Annual Meeting of the Association for Computational Linguistics (Volume 3: System Demonstrations)},
  address={Bangkok, Thailand},
  publisher={Association for Computational Linguistics},
  year={2024},
  url={http://arxiv.org/abs/2403.13372}
}

@article{huffman,
  title={Design and analysis of dynamic Huffman codes},
  author={Vitter, Jeffrey Scott},
  journal={Journal of the ACM (JACM)},
  volume={34},
  number={4},
  pages={825--845},
  year={1987},
  publisher={ACM New York, NY, USA}
}

@article{duda2013asymmetric,
  title={Asymmetric numeral systems: entropy coding combining speed of huffman coding with compression rate of arithmetic coding},
  author={Duda, Jarek},
  journal={arXiv preprint arXiv:1311.2540},
  year={2013}
}

@article{marpe2003context,
  title={Context-based adaptive binary arithmetic coding in the H. 264/AVC video compression standard},
  author={Marpe, Detlev and Schwarz, Heiko and Wiegand, Thomas},
  journal={IEEE Transactions on circuits and systems for video technology},
  volume={13},
  number={7},
  pages={620--636},
  year={2003},
  publisher={IEEE}
}

@article{q_coder_adaptive,
  title={An overview of the basic principles of the Q-coder adaptive binary arithmetic coder},
  author={Pennebaker, William B. and Mitchell, Joan L. and Langdon, Glen G and Arps, Ronald B},
  journal={IBM Journal of research and development},
  volume={32},
  number={6},
  pages={717--726},
  year={1988},
  publisher={IBM}
}

@article{huffman1952method,
  title={A method for the construction of minimum-redundancy codes},
  author={Huffman, David A},
  journal={Proceedings of the IRE},
  volume={40},
  number={9},
  pages={1098--1101},
  year={1952},
  publisher={IEEE}
}

@techreport{deutsch1996deflate,
  title={DEFLATE compressed data format specification version 1.3},
  author={Deutsch, Peter},
  year={1996}
}

@inproceedings{zafrir2019q8bert,
  title={Q8bert: Quantized 8bit bert},
  author={Zafrir, Ofir and Boudoukh, Guy and Izsak, Peter and Wasserblat, Moshe},
  booktitle={2019 Fifth Workshop on Energy Efficient Machine Learning and Cognitive Computing-NeurIPS Edition (EMC2-NIPS)},
  pages={36--39},
  year={2019},
  organization={IEEE}
}

@article{yao2022zeroquant,
  title={Zeroquant: Efficient and affordable post-training quantization for large-scale transformers},
  author={Yao, Zhewei and Yazdani Aminabadi, Reza and Zhang, Minjia and Wu, Xiaoxia and Li, Conglong and He, Yuxiong},
  journal={Advances in Neural Information Processing Systems},
  volume={35},
  pages={27168--27183},
  year={2022}
}

@article{dettmers2023spqr,
  title={Spqr: A sparse-quantized representation for near-lossless llm weight compression},
  author={Dettmers, Tim and Svirschevski, Ruslan and Egiazarian, Vage and Kuznedelev, Denis and Frantar, Elias and Ashkboos, Saleh and Borzunov, Alexander and Hoefler, Torsten and Alistarh, Dan},
  journal={arXiv preprint arXiv:2306.03078},
  year={2023}
}

@article{blalock2018sprintz,
  title={Sprintz: Time series compression for the internet of things},
  author={Blalock, Davis and Madden, Samuel and Guttag, John},
  journal={Proceedings of the ACM on Interactive, Mobile, Wearable and Ubiquitous Technologies},
  volume={2},
  number={3},
  pages={1--23},
  year={2018},
  publisher={ACM New York, NY, USA}
}

@inproceedings{burtscher2007high,
  title={High throughput compression of double-precision floating-point data},
  author={Burtscher, Martin and Ratanaworabhan, Paruj},
  booktitle={2007 Data Compression Conference (DCC'07)},
  pages={293--302},
  year={2007},
  organization={IEEE}
}

@article{liakos2022chimp,
  title={Chimp: efficient lossless floating point compression for time series databases},
  author={Liakos, Panagiotis and Papakonstantinopoulou, Katia and Kotidis, Yannis},
  journal={Proceedings of the VLDB Endowment},
  volume={15},
  number={11},
  pages={3058--3070},
  year={2022},
  publisher={VLDB Endowment}
}

@article{liu2021decomposed,
  title={Decomposed bounded floats for fast compression and queries},
  author={Liu, Chunwei and Jiang, Hao and Paparrizos, John and Elmore, Aaron J},
  journal={Proceedings of the VLDB Endowment},
  volume={14},
  number={11},
  pages={2586--2598},
  year={2021},
  publisher={VLDB Endowment}
}

@article{kuschewski2023btrblocks,
  title={Btrblocks: Efficient columnar compression for data lakes},
  author={Kuschewski, Maximilian and Sauerwein, David and Alhomssi, Adnan and Leis, Viktor},
  journal={Proceedings of the ACM on Management of Data},
  volume={1},
  number={2},
  pages={1--26},
  year={2023},
  publisher={ACM New York, NY, USA}
}

@article{vohra2016practical,
  title={Practical Hadoop Ecosystem},
  author={Vohra, Deepak},
  journal={Chapter in Apache Parquet},
  volume={177},
  pages={178},
  year={2016},
  publisher={Springer}
}

@article{diffenderfer2019error,
  title={Error analysis of ZFP compression for floating-point data},
  author={Diffenderfer, James and Fox, Alyson L and Hittinger, Jeffrey A and Sanders, Geoffrey and Lindstrom, Peter G},
  journal={SIAM Journal on Scientific Computing},
  volume={41},
  number={3},
  pages={A1867--A1898},
  year={2019},
  publisher={SIAM}
}

@article{lindstrom2014fixed,
  title={Fixed-rate compressed floating-point arrays},
  author={Lindstrom, Peter},
  journal={IEEE transactions on visualization and computer graphics},
  volume={20},
  number={12},
  pages={2674--2683},
  year={2014},
  publisher={IEEE}
}

@inproceedings{liang2018efficient,
  title={An efficient transformation scheme for lossy data compression with point-wise relative error bound},
  author={Liang, Xin and Di, Sheng and Tao, Dingwen and Chen, Zizhong and Cappello, Franck},
  booktitle={2018 IEEE International Conference on Cluster Computing (CLUSTER)},
  pages={179--189},
  year={2018},
  organization={IEEE}
}

@inproceedings{di2016fast,
  title={Fast error-bounded lossy HPC data compression with SZ},
  author={Di, Sheng and Cappello, Franck},
  booktitle={2016 ieee international parallel and distributed processing symposium (ipdps)},
  pages={730--739},
  year={2016},
  organization={IEEE}
}

@inproceedings{tao2017significantly,
  title={Significantly improving lossy compression for scientific data sets based on multidimensional prediction and error-controlled quantization},
  author={Tao, Dingwen and Di, Sheng and Chen, Zizhong and Cappello, Franck},
  booktitle={2017 IEEE International Parallel and Distributed Processing Symposium (IPDPS)},
  pages={1129--1139},
  year={2017},
  organization={IEEE}
}

@misc{git-lfs,
  author       = {GitHub},
  title        = {Git Large File Storage (LFS)},
  year         = {2024},
  howpublished = {\url{https://github.com/git-lfs/git-lfs}}
}

@techreport{netapp-tr3966,
  author       = {NetApp},
  title        = {NetApp ONTAP 9 Storage Efficiency Guide},
  institution  = {NetApp},
  year         = {2020},
  number       = {TR-3966},
  howpublished          = {\url{https://www.netapp.com/media/19753-tr-3966.pdf}}
}

@misc{dell-datadomain,
  author       = {{Dell Technologies}},
  title        = {Understanding Data Domain Compression},
  year         = {2023},
  howpublished          = {\url{https://www.dell.com/support/kbdoc/en-us/000003886/86266-understanding-datadomain-compression}}
}

@misc{dell-website,
  author       = {{Dell Technologies}},
  title        = {PowerProtect DD Backup Appliances},
  year         = {2024},
  howpublished          = {\url{https://www.dell.com/en-us/dt/data-protection/powerprotect-backup-dd-appliances/powerprotect-dd-backup-appliances.htm}}
}

@misc{netapp-ontap,
  author       = {{NetApp}},
  title        = {ONTAP Data Management Software},
  year         = {2024},
  howpublished          = {\url{https://www.netapp.com/data-management/ontap-data-management-software/}}
}

@misc{huggingface-lfs-analysis,
  author       = {{Xet Team}},
  title        = {{Git LFS Usage across the Hub}},
  year         = {2024},
  howpublished = {\url{https://huggingface.co/spaces/xet-team/lfs-analysis}}
}

@inproceedings{huang2012erasure,
  title={Erasure coding in windows azure storage},
  author={Huang, Cheng and Simitci, Huseyin and Xu, Yikang and Ogus, Aaron and Calder, Brad and Gopalan, Parikshit and Li, Jin and Yekhanin, Sergey},
  booktitle={2012 USENIX Annual Technical Conference (USENIX ATC 12)},
  pages={15--26},
  year={2012}
}

@misc{aws_s3,
  author       = {{Amazon Web Services}},
  title        = {Amazon S3: A Simple Storage Service},
  year         = {2006},
  howpublished = {\url{https://aws.amazon.com/s3/}}
}

@misc{s3_interview,
  title        = {{Building and operating a pretty big storage system called S3}},
  year         = {2023},
  howpublished = {\url{https://www.allthingsdistributed.com/2023/07/building-and-operating-a-pretty-big-storage-system.html}}
}

@misc{safetensors,
  author       = {{Hugging Face}},
  title        = {Safetensors Documentation},
  year         = {2024},
  howpublished          = {\url{https://huggingface.co/docs/safetensors/en/index}}
}

@misc{gguf,
  author       = {{ggml-org}},
  title        = {GGUF: GGML Universal File Format Specification},
  year         = {2024},
  howpublished = {\url{https://github.com/ggml-org/ggml/blob/master/docs/gguf.md}}
}

@misc{hdf_doc,
  title        = {{HDF Software Documentation}},
  howpublished = {\url{https://support.hdfgroup.org/documentation/index.html}}
}

@misc{meta-llama-3.1-8b,
  author       = {{Meta AI}},
  title        = {Llama 3.1 8B},
  year         = {2024},
  howpublished = {\url{https://huggingface.co/meta-llama/Llama-3.1-8B}}
}

@misc{wikipedia-vcdiff,
  author       = {{Wikipedia contributors}},
  title        = {VCDIFF -- Wikipedia{,} The Free Encyclopedia},
  year         = {2024},
  howpublished = {\url{https://en.wikipedia.org/wiki/VCDIFF}}
}

@misc{bsdiff,
  author       = {Colin Percival},
  title        = {bsdiff - Binary diff/patch utility},
  year         = {2003},
  howpublished          = {\url{https://www.daemonology.net/bsdiff/}}
}

@article{pelkonen2015gorilla,
  title={Gorilla: A fast, scalable, in-memory time series database},
  author={Pelkonen, Tuomas and Franklin, Scott and Teller, Justin and Cavallaro, Paul and Huang, Qi and Meza, Justin and Veeraraghavan, Kaushik},
  journal={Proceedings of the VLDB Endowment},
  volume={8},
  number={12},
  pages={1816--1827},
  year={2015},
  publisher={VLDB Endowment}
}

@misc{hf-model-cards,
  author       = {{Hugging Face}},
  title        = {Model Cards - Hugging Face Documentation},
  year         = {2024},
  howpublished = {\url{https://huggingface.co/docs/hub/en/model-cards}}
}

@inproceedings{dedup_tradeoff_fast15,
author = {Fu, Min and Feng, Dan and Hua, Yu and He, Xubin and Chen, Zuoning and Xia, Wen and Zhang, Yucheng and Tan, Yujuan},
title = {Design tradeoffs for data deduplication performance in backup workloads},
year = {2015},
isbn = {9781931971201},
publisher = {USENIX Association},
address = {USA},
booktitle = {Proceedings of the 13th USENIX Conference on File and Storage Technologies},
pages = {331–344},
numpages = {14},
location = {Santa Clara, CA},
series = {FAST'15}
}

@inproceedings{backup_dedup_fast12,
author = {Wallace, Grant and Douglis, Fred and Qian, Hangwei and Shilane, Philip and Smaldone, Stephen and Chamness, Mark and Hsu, Windsor},
title = {Characteristics of backup workloads in production systems},
year = {2012},
publisher = {USENIX Association},
address = {USA},
booktitle = {Proceedings of the 10th USENIX Conference on File and Storage Technologies},
pages = {4},
numpages = {1},
location = {San Jose, CA},
series = {FAST'12}
}

@misc{dong2024stbllmbreaking1bitbarrier,
      title={STBLLM: Breaking the 1-Bit Barrier with Structured Binary LLMs}, 
      author={Peijie Dong and Lujun Li and Yuedong Zhong and Dayou Du and Ruibo Fan and Yuhan Chen and Zhenheng Tang and Qiang Wang and Wei Xue and Yike Guo and Xiaowen Chu},
      year={2024},
      eprint={2408.01803},
      archivePrefix={arXiv},
      primaryClass={cs.LG},
      howpublished = {\url{https://arxiv.org/abs/2408.01803}}, 
}

@inproceedings{dedup_study_fast11,
author = {Meyer, Dutch T. and Bolosky, William J.},
title = {A study of practical deduplication},
year = {2011},
isbn = {9781931971829},
publisher = {USENIX Association},
address = {USA},
booktitle = {Proceedings of the 9th USENIX Conference on File and Stroage Technologies},
pages = {1},
numpages = {1},
location = {San Jose, California},
series = {FAST'11}
}

@article{metropolis1949monte,
  title={The monte carlo method},
  author={Metropolis, Nicholas and Ulam, Stanislaw},
  journal={Journal of the American statistical association},
  volume={44},
  number={247},
  pages={335--341},
  year={1949},
  publisher={Taylor \& Francis}
}

@misc{bartowski-gemma-gguf,
  author       = {{bartowski}},
  title        = {gemma-2-9b-it-GGUF},
  year         = {2024},
  howpublished = {\url{https://huggingface.co/bartowski/gemma-2-9b-it-GGUF}}
}

@article{yang2024qwen2,
  title={Qwen2. 5 technical report},
  author={Yang, An and Yang, Baosong and Zhang, Beichen and Hui, Binyuan and Zheng, Bo and Yu, Bowen and Li, Chengyuan and Liu, Dayiheng and Huang, Fei and Wei, Haoran and others},
  journal={arXiv preprint arXiv:2412.15115},
  year={2024}
}

@misc{meta2024llama3,
  author       = {{Meta AI}},
  title        = {Introducing Llama 3.1: Our most capable models to date},
  howpublished = {\url{https://ai.meta.com/blog/meta-llama-3-1/}},
  year         = {2024}
}

@misc{meta2024llama3p1,
  author       = {{Meta AI}},
  title        = {Introducing Meta Llama 3: The most capable openly available LLM to date},
  howpublished = {\url{https://ai.meta.com/blog/meta-llama-3/}},
  year         = {2024}
}

@misc{meta2025llama4,
  author       = {{Meta AI}},
  title        = {The Llama 4 herd: The beginning of a new era of natively multimodal AI innovation},
  howpublished = {\url{https://ai.meta.com/blog/llama-4-multimodal-intelligence/}},
  year         = {2025}
}

@article{gemma2024,
  title={Gemma 2: Improving Open Language Models at a Practical Size},
  author={Morgane Riviere and others},
  journal={arXiv preprint arXiv:2408.00118},
  year={2024},
  url={https://arxiv.org/abs/2408.00118}
}

@misc{meta2024llama3-2,
  author = {{Meta AI}},
  title = {Llama 3.2: Revolutionizing Edge AI and Vision with Open, Customizable Models},
  howpublished = {\url{https://ai.meta.com/blog/llama-3-2-connect-2024-vision-edge-mobile-devices/}},
  year         = {2024}
}

@inproceedings{frantar2023sparsegpt,
  title={Sparsegpt: Massive language models can be accurately pruned in one-shot},
  author={Frantar, Elias and Alistarh, Dan},
  booktitle={International Conference on Machine Learning},
  pages={10323--10337},
  year={2023},
  organization={PMLR}
}

@article{sun2023simple,
  title={A simple and effective pruning approach for large language models},
  author={Sun, Mingjie and Liu, Zhuang and Bair, Anna and Kolter, J Zico},
  journal={arXiv preprint arXiv:2306.11695},
  year={2023}
}

@misc{tensorflowhub,
  author       = {Google},
  title        = {{TensorFlow Hub: A Library for Reusable Machine Learning Modules}},
  year         = {2018},
  howpublished = {\url{https://www.tensorflow.org/hub}}
}

@misc{huggingface,
  author       = {Hugging Face},
  title        = {Hugging Face},
  howpublished = {\url{https://huggingface.co/}},
  year         = {2023}
}

@misc{hf_chunks,
  author       = {XetHub},
  title        = {From Chunks to Blocks: Accelerating Uploads and Downloads on the Hub},
  howpublished = {\url{https://huggingface.co/blog/from-chunks-to-blocks}},
  year         = {2025}
}

@inproceedings{yao2025deltazip,
  title={DeltaZip: Efficient Serving of Multiple Full-Model-Tuned LLMs},
  author={Yao, Xiaozhe and Hu, Qinghao and Klimovic, Ana},
  booktitle={Proceedings of the Twentieth European Conference on Computer Systems},
  pages={110--127},
  year={2025}
}

@article{liu2024bitdelta,
  title={Bitdelta: Your fine-tune may only be worth one bit},
  author={Liu, James and Xiao, Guangxuan and Li, Kai and Lee, Jason D and Han, Song and Dao, Tri and Cai, Tianle},
  journal={Advances in Neural Information Processing Systems},
  volume={37},
  pages={13579--13600},
  year={2024}
}

@article{yang2025qwen3,
  title={Qwen3 technical report},
  author={Yang, An and Li, Anfeng and Yang, Baosong and Zhang, Beichen and Hui, Binyuan and Zheng, Bo and Yu, Bowen and Gao, Chang and Huang, Chengen and Lv, Chenxu and others},
  journal={arXiv preprint arXiv:2505.09388},
  year={2025}
}

@article{team2025gemma,
  title={Gemma 3 technical report},
  author={Team, Gemma and Kamath, Aishwarya and Ferret, Johan and Pathak, Shreya and Vieillard, Nino and Merhej, Ramona and Perrin, Sarah and Matejovicova, Tatiana and Ram{\'e}, Alexandre and Rivi{\`e}re, Morgane and others},
  journal={arXiv preprint arXiv:2503.19786},
  year={2025}
}

@misc{Qwen-math,
  author       = {{Qwen}},
  title        = {{Introducing Qwen2-Math}},
  year         = {2024},
  howpublished = {\url{https://qwenlm.github.io/blog/qwen2-math/}},
}

@misc{Qwen-coder,
  author       = {{Qwen}},
  title        = {{Qwen2.5-Coder: Code More, Learn More!}},
  year         = {2024},
  howpublished = {\url{https://qwen.ai/blog?id=d9c66f64e7a2e156790c7991df3c803a7c3f96cd&from=research.research-list}},
}

@misc{Qwen-VL,
  author       = {{Qwen}},
  title        = {{Qwen2.5 VL! Qwen2.5 VL! Qwen2.5 VL!}},
  year         = {2025},
  howpublished = {\url{https://qwen.ai/blog?id=c5e7415d9a9e89adc18c59d9e466e5a1a459b8f4&from=research.research-list}},
}
